\documentclass{article}

\usepackage{amsfonts}
\usepackage{amssymb}
\usepackage{amsthm}
\usepackage{amsmath}

\usepackage{graphicx}

\usepackage{subfigure}
\usepackage{bm}             %% boldmath-command: \bm{}
\usepackage[mathscr]{eucal}
\usepackage{cancel}
\usepackage{wasysym}

\usepackage{oubraces}

\def\pmb#1{\setbox0=\hbox{$#1$}%
  \kern-.025em\copy0\kern-\wd0
  \kern.05em\copy0\kern-\wd0
  \kern-.025em\raise.0433em\box0}
\def\pmbs#1{\setbox0=\hbox{$\scriptstyle #1$}%
  \kern-.0175em\copy0\kern-\wd0
  \kern.035em\copy0\kern-\wd0
  \kern-.0175em\raise.0303em\box0}

\def\be{\begin{equation}}
\def\ee{\end{equation}}
\def\bea{\begin{eqnarray}}
\def\eea{\end{eqnarray}}

\def\cn{{\cal N}}

\def\cT{{\cal T}}

\newcommand{\lin}{l_{\mathrm{in}}}
\newcommand{\lout}{l_{\mathrm{out}}}

\def\parb{\pmb{\partial}}

\def\la{\langle}
\def\ra{\rangle}

\def\hsp5{\hspace{5mm}}

\newcommand{\sfrac}[2]{\textstyle{\frac{#1}{#2}}}
\newcommand{\textfrac}[2]{{\textstyle{\frac{#1}{#2}}}}
\def\case#1/#2{\textstyle\frac{#1}{#2}}

\newcommand{\EE}[2]{E^{#1}_{\:\: #2}}

\renewcommand{\vector}[1]{\bm{#1}}

\renewcommand{\u}{\mathsf{u}}

\newcommand{\kh}{K}
\newcommand{\khl}{L}
\newcommand{\khp}{\bar{K}}
\newcommand{\khpl}{\bar{L}}

\newcommand{\era}{\overline{\mathrm{era}}}

\newcommand{\Thi}{\mathcal{T}_{\mathsf{Hi}}}

\newcommand{\Tco}{\mathcal{T}_{\mathsf{Jo}}}

\theoremstyle{plain}

\theoremstyle{remark}

\begin{document}

%%%%%%%%%%%%%%%%%%%%%%%%%%%%%%%%%%%%%%%%%%%%%%%%%%%%%%%%%%%%%%%%%%%
\title{\sc Recent developments concerning generic spacelike singularities}
%%%%%%%%%%%%%%%%%%%%%%%%%%%%%%%%%%%%%%%%%%%%%%%%%%%%%%%%%%%%%%%%%%%

\author{\sc Claes Uggla\thanks{Electronic address: {\tt claes.uggla@kau.se}} \\
{\small\em Department of Physics, University of Karlstad,}\\
{\small\em S-65188 Karlstad, Sweden}}

%%%%%%%%%%%%%%%%%%%%%%%%%%%%%%%%%%%%%%%%%%%%%%%%%%%%%%%%%%%%%%%%%%%
%\date{\normalsize{April 25, 2013}}
\date{}
%%%%%%%%%%%%%%%%%%%%%%%%%%%%%%%%%%%%%%%%%%%%%%%%%%%%%%%%%%%%%%%%%%%
\maketitle
%%%%%%%%%%%%%%%%%%%%%%%%%%%%%%%%%%%%%%%%%%%%%%%%%%%%%%%%%%%%%%%%%%%

%%%%%%%%%%%%%%%%%%%%%%%%%%%%%%%%%%%%%%%%%%%%%%%%%%%%%%%%%%%%%%%%%%%
\begin{abstract}
%%%%%%%%%%%%%%%%%%%%%%%%%%%%%%%%%%%%%%%%%%%%%%%%%%%%%%%%%%%%%%%%%%%

This is a review of recent progress concerning generic spacelike
singularities in general relativity. For brevity the main focus is on
singularities in vacuum spacetimes, although the connection with, and the
role of, matter for generic singularity formation is also commented on. The
paper describes recent developments in two areas and show how these are
connected within the context of the conformally Hubble-normalized state space
approach. The first area is oscillatory singularities in spatially
homogeneous cosmology and the connection between asymptotic behaviour and
heteroclinic chains. The second area concerns oscillatory singularities in
inhomogeneous models, especially spike chains and recurring spikes. The
review also outlines some underlying reasons for why the structures that are
the foundation for generic oscillatory behaviour exists at all, which entails
discussing how underlying physical principles and applications of solution
generating techniques yield hierarchical structures and connections between
them. Finally, it is pointed out that recent progress concerning generic
singularities motivates some speculations that suggest that a paradigm shift
concerning their physical role, and what mathematical issues to address,
might be in order.

%%%%%%%%%%%%%%%%%%%%%%%%%%%%%%%%%%%%%%%%%%%%%%%%%%%%%%%%%%%%%%%%%%
\end{abstract}
%%%%%%%%%%%%%%%%%%%%%%%%%%%%%%%%%%%%%%%%%%%%%%%%%%%%%%%%%%%%%%%%%%%
\centerline{\bigskip\noindent PACS number(s):
04.20.Dw, 04.20.Ha, 04.20.-q, 05.45.-a, 98.80.Jk\hfill %{gr-qc/yymmnnn}...
}\vfill

%%%%%%%%%%%%%%%%%%%%%%%%%%%%%%%%%%%%%%%%%%%%%%%%%%%%%%%%%%%%%%%%%%
\section{Introduction}
%%%%%%%%%%%%%%%%%%%%%%%%%%%%%%%%%%%%%%%%%%%%%%%%%%%%%%%%%%%%%%%%%%

This review describes recent results about generic spacelike singularities,
but this is not its only purpose; the goal is to also provide a mathematical
and physical context to these developments. This context will, at least
partly, explain why one has been able to obtain any results at all, but it
also suggests that future progress to some extent requires revising some
cherished concepts, as well as challenging some widely held beliefs, values
and goals. To provide a reasonably short and accessible overall picture, it
is necessary to make some restrictions, which are chosen as follows: (i)
4-dimensional General Relativity (GR). (ii) Vacuum. (iii) Generic oscillatory
spacelike singularities. (iv) Focus on main ideas in order to provide an
overall picture. To accomplish this, we refer to the literature for some of
the details while other details and open issues, of different levels of
difficulty, are yet to be worked out and solved, although outlines for how to
do this will sometimes be given.

Any investigation about the detailed nature of generic spacelike
singularities has to take into account the work by Belinski\v{\i},
Khalatnikov and Lifshitz (BKL)~\cite{lk63,bkl70,bkl82}. This work began by
considering Einstein's field equations in synchronous coordinates and by
dropping all spatial derivatives, which geometrically corresponds to
neglecting the Ricci 3-curvature of the spatial surfaces of the synchronous
coordinate system, as well as all matter terms~\cite{lk63}. This procedure
leads to a set of ordinary differential equations (ODEs) that are identical
to those obtained in the vacuum case by imposing spatial homogeneity and an
associated simply transitive Abelian symmetry group, which results in the
vacuum Bianchi type I models whose solution is the well-known Kasner
solution. But in the general inhomogeneous context the constants of
integration that appear in the Kasner solution are replaced by spatially
dependent functions, leading to a "generalized Kasner solution," even though
it is not a solution to Einstein's field equations at all. Instead the
relevance of the generalized Kasner solution is as a building block when one
attempts to construct generic asymptotic solutions; however, the generalized
Kasner solution itself does not have a sufficient number of spatial functions
to be a generic solution, something that originally led Lifshitz and
Khalatnikov to conclude that singularities are not a generic feature of
GR~\cite{lk63}.

This state of affairs changed, however, when the singularity theorem of
Penrose appeared in 1965~\cite{pen65}. According to this theorem, and later
related ones, singularities occur generically under quite general
circumstances. But singularity theorems of `Penrose type' say little about
the nature of generic singularities, and it should also be pointed out that
there are generic spacetimes without singularities.\footnote{This is
illustrated by the global stability of the Minkowski
spacetime~\cite{chrkla93}.} Prompted by this state of affairs, BKL noted that
the generalized Kasner `solutions' are unstable, which heuristically follows
from inserting them into the terms of the Einstein field equations that
drives the evolution and studying these terms temporal behaviour. Further
insights were gained from considerations of spatially homogeneous (SH)
models, especially Bianchi type IX. For these latter models they found that
generic solutions could be heuristically approximated asymptotically by
piecewise joining different Kasner states by means of different Bianchi type
II vacuum solutions in a manner that resulted in infinite Kasner
oscillations, a result that was also reached independently by Misner by means
of Hamiltonian methods~\cite{mis69a,mis69b}. Moreover, the Bianchi type II
solutions allowed for a discretization of the dynamics in terms of sequences
of Kasner states by means of a map~\cite{bkl70,bkl82,khaetal85} that turned
out to be connected with chaotic properties, something that has inspired many
subsequent papers, see e.g.~\cite{heiugg09a,ugg13a} for references.

Coming back to the general inhomogeneous case, armed with the insights from
the SH case, and by studying the effects of inserting the generalized Kasner
solution into the terms of the Einstein field equations that drives the
evolution, BKL claimed that certain terms in the Einstein field equations
could be dropped towards a generic spacelike singularity. In particular, this
held for the perfect fluid terms in the Einstein field equations if the
equations of state was sufficiently soft, e.g. dust or radiation. Hence,
according to BKL, "matter does not matter" for the spacetime geometry in the
vicinity of a generic spacelike singularity, even though the energy density
blows up, which makes it natural to start investigations about generic
spacelike singularities by considering the vacuum equations, as is done here.
Furthermore, by dropping terms BKL obtained effective ODEs that resulted in
`generalized' Bianchi type II vacuum `solutions' that temporally joined
different generalized Kasner states (in analogy with the type IX case) along
\emph{individual} timelines~\cite{bkl82}.\footnote{In addition BKL linearly
perturbed the generalized Bianchi type I and II `solutions' and obtained new
effective ODEs, which resulted in statements about `rotation and freezing of
Kasner axes'~\cite{bkl82}.} The resulting BKL picture for the `\emph{vacuum
dominated}' case can be summed up in the following `\emph{locality
conjecture}': The dynamics towards a generic spacelike singularity for
general \emph{inhomogeneous} models can be described as being \emph{`local},'
in the sense that each spatial point is assumed to evolve towards the
singularity individually and independently of its neighbors with its
evolution described by a system of ODEs; moreover, asymptotically the
evolution along individual timelines is described by a sequence of Bianchi
type II solutions that connect different Kasner states in an oscillatory
manner.

The locality conjecture leads to the question: Why should generic solutions
behave in this way? Presumably it has to do with asymptotic causal features.
Causal properties constitute an important feature in GR, notably in the
derivation of the singularity theorems of Penrose and
Hawking~\cite{pen65,hawpen70}. A generic spacelike singularity is expected to
be a scalar curvature singularity, since a generic singularity presumably is
associated with ultra-strong gravity and since a non-scalar curvature
singularity requires fine tuning. Moreover, it seems reasonable that
increasing ultra-strong gravity may lead to the formation of particle
horizons that shrink to zero size in all directions along any time line that
approaches the singularity, thus increasingly prohibiting communication, a
phenomenon that was called \emph{asymptotic silence}
in~\cite{uggetal03,heietal09,ugg13a,limetal06}. This would then `explain' why
asymptotic locality happens, but unfortunately things do not seem to be that
simple. Presumably asymptotic silence is a necessary requirement for
asymptotic locality, but it is not sufficient. As we will see, generic
singularities are not only associated with asymptotic locality, but also with
\emph{recurring spikes}, \emph{non-local} evolution along certain timelines
that is described by PDEs~\cite{andetal05,lim04,heietal12}, and it seems that
this type of behaviour is also associated with asymptotic silence (i.e.,
ultra-strong gravity does not always lead to asymptotic locality, nor to
asymptotic silence as illustrated by null singularities); for various
examples of asymptotic silence and asymptotic silence-breaking,
see~\cite{limetal06}.

Although we will focus on vacuum in this paper, it is worth pointing out that
causal properties of matter also seem to be important for if and how matter
influences the spacetime geometry in the vicinity of a generic spacelike
singularity. Towards such a singularity there is a competition between how
matter and gravity nonlinearly generate spacetime curvature. Examples suggest
that "matter does not matter" for matter sources with characteristics with
speed less than the speed of light; such models are therefore said to be
\emph{asymptotically vacuum dominated\/}. But if the matter equations have
characteristics with a speed that is equal to the speed of light, then the
tug of war between matter and pure gravity generating gravity results in a
draw; the Ricci and Weyl scalars obtain generic asymptotic amplitude
magnitudes of the same order. Moreover, in this case there exist interesting
connections between spin and spacetime curvature in the vicinity of a generic
spacelike singularity. A massless scalar field leads to asymptotic locality
and convergence to specific limits, a phenomenon that can be characterized as
\emph{asymptotic local self-similarity}. An electromagnetic field on the
other hand leads to non-convergent infinite oscillatory behaviour. Finally, a
perfect fluid, for which the equation of state in the infinite energy density
limit leads to a speed of sound that is greater than that of light, yields an
asymptotic locally self-similar and isotropic singularity for which the Ricci
scalar dominates over the Weyl scalar~\cite{sanugg10,heisan12}. These
tantalizing hints, that ultra-strong gravity seems to be connected with some
of the main properties of matter, is a fairly unexplored area of research
that deserves further study, and it is also of interest to also go beyond GR
in this context, as exemplified by the work of Damour {\it et
al\/}~\cite{dametal03}.

This latter work makes use of Hamiltonian methods that were developed by
Chitr\'e and Misner for Bianchi type IX~\cite{chi72,grav73}. It is worth
pointing out that this heuristic approach, which is closely connected to that
of BKL, gives rise to asymptotic constants of the motion, which are difficult
to obtain by other methods. Moreover, these methods have revealed a
remarkable connections between asymptotic dynamics of generic spacelike
singularities and Kac-Moody algebras~\cite{dametal03}.

Finally, the BKL picture, although likely to describe many asymptotic
features of generic spacelike singularities, contains several highly
non-trivial assumptions. Moreover, consistency arguments for the remarkably
simple BKL picture do not exclude other behaviour as regards singularities,
special or generic. Indeed, during the last few years there has been a number
of developments, analytical, numerical, and heuristic, that all point at that
the BKL scenario is part of a bigger startlingly subtle picture, with a web
of intriguing hierarchical structures, which furthermore hints at that you
have to go beyond BKL in order to fully understand the BKL picture.

The outline of the paper is as follows. The next section describes the
conformally `Hubble-normalized' asymptotically regularized state space
picture~\cite{rohugg05,heietal09}. This framework leads to a new context for
models with symmetries, which makes it possible to provide a rigorous and
simple description of asymptotic locality and the BKL picture; moreover, this
state space picture also makes it possible to describe the structures that
give rise to non-local recurring spikes. Section~\ref{Sec:hier} describes and
discusses the hierarchy of invariant subsets in the Hubble-normalized state
space picture that is the foundation for describing and understanding generic
spacelike singularities. Section~\ref{Sec:concat} outlines how the BKL
picture can be transformed to a rigorous description within the context of
the Hubble-normalized state space picture in terms of attractors,
transitions, concatenation and heteroclinic chains. Furthermore, the
gauge-invariant discrete representation of these structures in terms of
Kasner sequences naturally leads to a description of recent progress
concerning generic asymptotics in Bianchi types VIII and IX, which via the
Hubble-normalized state space picture becomes tied to the generic BKL
picture. Section~\ref{Sec:spikes} goes beyond the BKL picture and outlines
recent progress concerning spike chains and recurring spikes. In addition a
comparison of some statistical properties that arise from asymptotic BKL and
spike dynamics, respectively, is given. The paper concludes with a discussion
about the physical role of generic singularities, and the dangers and
possibilities of special models in contexts such as cosmic censorship.

%Section~\ref{Sec:beyondBKL1} describe recent results that pertains and partly
%explains an elaborated BKL picture. To do so concepts such as concatenation
%and heteroclinic chains, which gives rigor to Kasner oscillations, are
%introduced. Section~\ref{Sec:beyondBKL2} discusses recent developments
%concerning non-BKL recurring spike formation, and the underlying mechanisms
%that provide the building blocks for describing such structures. The previous
%sections are mainly kinematical in content. In Section~\ref{Sec:dyn} we
%discuss the dynamical relevance of these structures. The paper is finally
%concluded with Section~\ref{Sec:concl} with some speculations about a need
%for revising certain concepts, goals and beliefs, as well as outlining some
%possible future fruitful goals and developments.

%%%%%%%%%%%%%%%%%%%%%%%%%%%%%%%%%%%%%%%%%%%%%%%%%%%%%%%%%%%%%%%%%%
\section{The conformally Hubble-normalized state space formulation}\label{Sec:confHubble}
%%%%%%%%%%%%%%%%%%%%%%%%%%%%%%%%%%%%%%%%%%%%%%%%%%%%%%%%%%%%%%%%%%

%-----------------------------------------------------------------
%\subsection{State space and field equations}\label{Subsec:Hubblestate}
%-----------------------------------------------------------------

Currently there are no theorems about generic inhomogeneous oscillatory
spacelike singularities. In this paper it is argued that it is a reasonable
proximate goal to construct an asymptotic solution in a small spatiotemporal
neighborhood in the vicinity of part of a generic inhomogeneous singularity,
and to postpone ultimate goals concerning issues such as cosmic censorship.
With the proximate goal in mind:
\begin{itemize}
\item Consider a 4-dimensional manifold $M$ endowed with a metric
    $\mathbf{g}$ of signature $(-+++)$ (the `physical spacetime') that
    satisfies Einstein's vacuum equations. By convention set $c = 1$,
    where $c$ is the speed of light.
\item Due to the present cultural prominence of BKL, the singularity is
    chosen to be located in the past. Assume that there exists a small
    spatiotemporal neighborhood of the singularity that can be foliated
    with spacelike leaves that asymptotically coincide with the
    singularity.
\item Assume that the expansion $\theta$ of the timelike future directed
    unit normal vector field $\mathbf{u}$ of the chosen foliation is
    positive in the small neighborhood and that $\theta \rightarrow
    \infty$ towards the singularity, i.e., the generic singularity is
    assumed to be a crushing singularity.
\item Use the characteristic scale of the problem associated with the
    expansion $\theta$, or rather, due to historical reasons, the Hubble
    variable $H=\frac{1}{3}\theta$, to \emph{conformally blow up} the
    small neighborhood, as follows:
    \begin{equation}
    {\bf G} = H^{2}{\bf g},
    \end{equation}
    where the unphysical metric ${\bf G}$ is dimensionless since ${\bf
    g}$ has dimension (length)$^2$, or, equivalently, (time)$^2$, and $H$
    has dimension (length)$^{-1}$.
\item Introduce an orthonormal frame of ${\bf G}$,
    $\{\boldsymbol{\Omega}^a = E^a\!_\mu dx^\mu\}$, $a=0,1,2,3,\, \mu =
    0,1,2,3$ (alternatively interpreted as a conformal orthonormal frame
    of ${\bf g}$), according to:
\begin{equation} \label{defconfon}
\mathbf{G} =  \eta_{a b}\, \boldsymbol{\Omega}^a \boldsymbol{\Omega}^b =
H^2 \eta_{a b} \boldsymbol{\omega}^a \boldsymbol{\omega}^b = H^2 \mathbf{g},
\end{equation}
where $\eta_{ab} = {\rm diag}[-1,1,1,1]$, and where
$\{\boldsymbol{\omega}^a = e^a\!_\mu dx^\mu\}$ is the associated
orthonormal frame of ${\bf g}$.
\item Adapt the frame to $\mathbf{u}$ by choosing a set of dual frame
    vectors $\parb_a = E_a\!^\mu \partial_{x^\mu} = H^{-1}{\bf e}_a =
    H^{-1}e_a\!^\mu \partial_{x^\mu}$, where ${\bf e}_a$ ($\parb_a$) are
    the frame vectors dual to $\boldsymbol{\omega}^a$
    ($\boldsymbol{\Omega}^a$) ($e_a\!^\mu e^b\!_\mu = \delta_a\!^b$,
    $E_a\!^\mu E^b\!_\mu = \delta_a\!^b$), by setting ${\bf e}_0 =
    \mathbf{u}$. Furthermore, choose ${\bf e}_0 = \mathbf{u}$ ($\parb_0 =
    H^{-1}\mathbf{u}$) to be tangential to the timelines, i.e., $\parb_0
    = (HN)^{-1}\partial_{x^0} =: {\cal N}^{-1}\partial_{x^0}$ (i.e. set
    the shift vector to zero), where $N$ is the lapse function and ${\cal
    N}$ the conformal lapse function. Hence $\mathbf{G} = \eta_{a b}\,
    \boldsymbol{\Omega}^a \boldsymbol{\Omega}^b = -\cn^2 (d x^0)^2 +
    \delta_{\alpha \beta} \EE{\alpha}{i} \EE{\beta}{j} dx^i dx^j = -\cn^2
    (d x^0)^2 + G_{ij} dx^i dx^j$, $\alpha = 1,2,3$, $i = 1,2,3$.
\item Introduce a set of variables given by the inverse Hubble variable
    $H^{-1}$, the Hubble-normalized frame variables $\cn,\, E_\alpha^i$,
    and the Hubble-normalized commutator (equivalently, connection)
    variables $q,\, \Sigma_{\alpha\beta},\, R_\alpha,\, N^{\alpha\beta},
    A_\alpha$, defined by the commutators
\begin{subequations}\label{commuta}
\begin{equation}
[\,\parb_{0}, \parb_{\alpha}\,] =
(\parb_{\alpha}\log \cn)\,\parb_{0} +
F_\alpha{}^\beta\,\parb_{\beta},\qquad
[\,\parb_{\alpha}, \parb_{\beta}\,] =
(2A_{[\alpha}\,\delta_{\beta]}{}^{\gamma} +
\epsilon_{\alpha\beta\delta}\,N^{\delta\gamma})\,\parb_{\gamma},
\end{equation}
where $[...]$ corresponds to anti-symmetrization and
\begin{equation}
F_\alpha{}^\beta := q\,\delta_{\alpha}{}^{\beta} -
\Sigma_{\alpha}{}^{\beta} -
\epsilon_{\alpha}{}^{\beta}{}_{\gamma}\,R^{\gamma},
\end{equation}
\end{subequations}
where $-q$, $\Sigma_{\alpha\beta}$ are the conformally Hubble-normalized
Hubble variable and the trace-free shear of $\parb_0$, respectively,
where $q$ also happens to be the physical deceleration parameter;
$R_\alpha$ is the Fermi rotation, which describes how the frame rotates
with respect to a Fermi propagated frame; $N^{\alpha\beta}$ and
$A_\alpha$ are spatial commutator functions that describe the 3-curvature
of the spatial part of the metric $\mathbf{G}$.
\item Impose the commutator equations, Jacobi identities and Einstein
    field equations on $H^{-1}$ and the conformally Hubble-normalized
    frame and commutator variables.
\end{itemize}

The quantities $\cn$ and $R_\alpha$ represent remaining gauge
freedom\footnote{There is also gauge freedom associated with changes of
spatial coordinates, which affects $E_\alpha{}^i$, however, since this only
`passively' enters the discussion about evolution we focus on here, we will
not be concerned with this freedom further in this paper.} since: (i) The
above assumptions and choices do not uniquely fix the foliation and there is
therefore remaining freedom in choosing $N$ and hence also $\cn$ (for some
analytic purposes there is no need to impose any explicit restriction on
$\cn$, and hence we do not do so below, but $\cn$ needs to be specified for
numerical investigations as well as for describing asymptotic non-local
phenomena such as recurring spikes). (ii) The quantity $R_\alpha$ can be
chosen freely since the spatial frame is only determined up to arbitrary
rotations, however, to obtain a deterministic system of equations $R_\alpha$
needs to be specified; below we will use so-called Fermi and Iwasawa frames,
although other choices will be discussed as well. Keeping the choice of $\cn$
and $R_\alpha$ open leads to the following (not fully gauge reduced)
\emph{state space}:
\begin{equation}\label{Xstatesp}
\bm{X}  = \left(E_{\alpha}{}^{i}, H^{-1}, \Sigma_{\alpha\beta}, A_\alpha, N_{\alpha\beta}\right) =
(E_{\alpha}{}^{i})\oplus H^{-1}\oplus\bm{S}, \quad\text{where}\quad
\bm{S}=\left(\Sigma_{\alpha\beta},A_\alpha,N_{\alpha\beta}\right)\!.
\end{equation}
The commutator equations, Jacobi identities and Einstein field equations
yield the following equations for the state space variables (where the
choices of $\cn$ and $R_\alpha$ are left unspecified):

\textit{Evolution equations}:
\begin{subequations}\label{alleveq}
\begin{alignat}{2}
\parb_{0}E_{\alpha}{}^{i} & =
F_{\alpha}{}^{\beta} E_{\beta}{}^{i}, & & \label{dl13comts}\\
\parb_{0} H^{-1} &= (1 + q)H^{-1}, & & \label{Heq}\\
\parb_{0}\Sigma_{\alpha\beta} &= -(2-q)\Sigma_{\alpha\beta} -
2\epsilon^{\gamma\delta}{}_{\langle\alpha}\,\Sigma_{\beta\rangle\gamma}\,R_{\delta}
- {}^{3}\!{S}_{\alpha\beta} &\quad + &\quad (I_\Sigma)_{\alpha\beta}, \label{dlsigdot}\\
\parb_{0}N^{\alpha\beta} & =
(3q\,\delta_{\gamma}{}^{(\alpha}-2F_{\gamma}{}^{(\alpha}{})\,N^{{\beta}){\gamma}}
 &\quad + &\quad (I_N)^{\alpha\beta},\label{dlndot}\\
\parb_{0}A_{\alpha} & =
F_{\alpha}{}^{\beta}A_{\beta}  &\quad + &\quad (I_A)_\alpha . \label{dladot}
\end{alignat}
\end{subequations}

\textit{Constraint equations}:
\begin{subequations}\label{allconeq}
\begin{alignat}{2}
\label{dl13com} 0 & = (\epsilon_{\alpha}{}^{\gamma\beta}\,
(\parb_{\gamma}-A_{\gamma}) - N_{\alpha}{}^{\beta})E_{\beta}{}^{i}, && \\
\label{dljacobi2} 0 & = &\quad &\quad (I_{J\omega})_\alpha ,\\
\label{dlgauss} 0 & = 1 - \Sigma^2 - \Omega_k &\quad - &\quad (I_G),\\
\label{dlcodazzi} 0 & =  -3A_{\beta}\Sigma_{\alpha}{}^{\beta}+
\epsilon_{\alpha}{}^{\beta\gamma}\Sigma_{\beta}{}^{\delta}N_{\delta\gamma}
&\quad + &\quad
(I_C)_\alpha ,\\
\label{dljacobi1} 0 & =  A_{\beta} N_{\alpha}{}^{\beta} &\quad + &\quad
(I_J)_\alpha ,
\end{alignat}
\end{subequations}
where
\begin{subequations}\label{defofmuch}
\begin{alignat}{3}
\Sigma^2 &:= \textfrac{1}{6}\Sigma_{\alpha\beta}\Sigma^{\alpha\beta},\quad &
q &= 2\Sigma^2 &\quad + &\quad (I_q),\\
\Omega_k &:= -\textfrac{1}{6}\,{}^{3}{\cal R},\qquad & {}^{3}\!{\cal R} &=
- 6A_{\alpha}A^\alpha - \textfrac{1}{2}B_\alpha{}^\alpha &\quad + &\quad (I_R),\\
B_{\alpha\beta} &:=  2N_{\alpha}{}^\gamma\,N_{\gamma\beta} -
N_\gamma{}^\gamma\,N_{\alpha\beta},\qquad & {}^{3}{\cal S}_{\alpha\beta} &=
B_{\langle\alpha\beta\rangle} -
2\epsilon^{\gamma\delta}{}_{\langle\alpha}\,N_{\beta\rangle\gamma}\,A_{\delta}
&\quad + &\quad (I_S)_{\alpha\beta} .
\end{alignat}
\end{subequations}
In the above equations $(\ldots)$ and $\la \ldots\ra$ denote spatial
symmetrization and trace-free symmetrization, respectively; ${}^{3}\!{\cal
R}, {}^{3}\!{\cal S}_{\alpha\beta}$ are the conformal scalar 3-curvature and
trace-free Ricci 3-curvature, respectively; for brevity we omit the
expressions for $(I_{\ast})_{**}$, which are given
in~\cite{heietal09,rohugg05}, but all involve terms of the type
$\parb_\gamma(\log {\cal N}, \log H^{-1}, \bm{S}, R_\alpha)$, and if these
quantities are zero so are all $(I_{\ast})_{**}$.

%%%%%%%%%%%%%%%%%%%%%%%%%%%%%%%%%%%%%%%%%%%%%%%%%%%%%%%%%%%%%%%%%%
\section{Invariant subset hierarchies}\label{Sec:hier}
%%%%%%%%%%%%%%%%%%%%%%%%%%%%%%%%%%%%%%%%%%%%%%%%%%%%%%%%%%%%%%%%%%

%-----------------------------------------------------------------
\subsection{Symmetry based invariant subset hierarchies}\label{Subsec:hierstate}
%-----------------------------------------------------------------

Due to that Killing vector fields are Ricci collinations, i.e., the Lie
derivative of the Ricci tensor with respect to a Killing vector is zero, such
symmetries are compatible with Einstein's (vacuum) equations and hence lead
to invariant subsets of those equations. This is also true for homothetic
Killing vectors, but not for proper conformal Killing vector fields, which
explains the rarity of solutions with such symmetries. In the present context
the most interesting isometry groups are those that act on spacelike orbits.
Imposing conditions on the Killing vectors and the algebras they form result
in a hierarchy of invariants subsets that play a key role for asymptotic
dynamics. This is especially true for the SH case, whose system of equations
can be derived straightforwardly from the conformally Hubble-normalized
picture.

%-----------------------------------------------------------------
\subsubsection*{The SH case in the conformally Hubble-normalized state space picture}\label{Subsec:SHstate}
%-----------------------------------------------------------------

To obtain the SH case in the conformally Hubble-normalized picture, choose a
symmetry compatible frame $\{\vector{e}_a\}$ such that $\vector{e}_0 =
N^{-1}\partial_{x^0}$ is orthogonal to the spatial symmetry surfaces, and
hence that $\vector{e}_\alpha = e_\alpha{}^i\partial_{x^i}$ are tangential to
the symmetry surfaces, and similarly for $\{\parb_a\}$. As a consequence the
lapse $N$ and the coordinate scalars $(H^{-1}, \bm{S}, R_\alpha)$, and hence
also ${\cal N}$, are functions of time only, and therefore $\parb_\gamma(
{\cal N}, H^{-1}, \bm{S}, R_\alpha) = 0$ (recall that $\parb_\gamma =
E_\gamma\!^i\partial_{x^i}$), which leads to that all $(I_{\ast})_{**}$
become zero. As a consequence, the evolution equation~\eqref{dl13comts} for
$E_\alpha{}^i$ decouple from the other equations, leaving a system for
$H^{-1}\oplus\bm{S}$, where the equation~\eqref{Heq} for $H^{-1}$ also
decouples, which leads to a \emph{reduced coupled dimensionless system} for
$\bm{S}$, given by the evolution
equations~\eqref{dlsigdot},~\eqref{dlndot},~\eqref{dladot} and the
constraints~\eqref{dlgauss},~\eqref{dlcodazzi},~\eqref{dljacobi1}. Since $q =
2\Sigma^2$, it follows from~\eqref{Heq} that $H^{-1}>0$ is monotonically
decreasing towards the past towards zero (see, e.g.~\cite{heiugg09a}), which
leads to a crushing singularity that is a scalar curvature singularity in the
generic case, although there are special cases that result in e.g. a
coordinate singularity.

If we in addition add general matter terms to the Einstein field equations,
as done in~\cite{rohugg05,sanugg10}, we obtain the general Einstein field
equations for the SH case, although these equations have to be complemented
with equations for the matter degrees of freedom. Specializing to (i) an
orthogonal (`non-tilted') perfect fluid (i.e. the fluid 4-velocity is given
by $\vector{e}_0$), and a linear equation of state, (ii) the SH class A
invariant subset $A_\alpha = 0$, (iii) and choose a frame for which
$N^{\alpha\beta}$ and $\Sigma_{\alpha\beta}$ are both diagonal, which is
possible for this particular set of SH models,\footnote{This is possible
locally; global topological issues may change the situation, as discussed
in~\cite{ashsam91}.} leads to the dynamical system introduced by Wainwright
and Hsu~\cite{waihsu89}, i.e., that approach is a special case of the present
geometrical framework.

%-----------------------------------------------------------------
\subsubsection*{The $G_2$ case in the conformally Hubble-normalized state space picture}\label{Subsec:G2state}
%-----------------------------------------------------------------

Models with two commuting spacelike Killing vectors, the so-called $G_2$
models, turn out to be important for asymptotic dynamics, moreover, they are
the simplest inhomogeneous vacuum models that (conjecturally) admit
oscillatory singularities. These models can be expressed in terms of a
general so-called spatial Iwasawa frame. Within the conformally
Hubble-normalized framework, such a spatial frame can be implemented by
setting
\begin{equation}
E_1\!^2 = E_1\!^3 = E_2\!^3 =0 \quad \Rightarrow \quad N_3=0, \quad
(R_1,R_2,R_3) = (-\Sigma_{23},\Sigma_{31},-\Sigma_{12}),
\end{equation}
where in the $G_2$ case one can choose local coordinates and a symmetry
adapted Iwasawa frame so that all variables depend on $x^0$ and $x^3$ only,
which in addition gives $N_{31}=N_{23}=N_{22}=A_1=A_2=0$, and where one of
the Codazzi constraints can be used to also impose $R_2=0=\Sigma_{31}$.
Moreover, due to the imposed symmetries, only $\parb_0$ and $\parb_3$ occur
in the Einstein field equations and the Jacobi identities. As a consequence
the evolution equations for all the remaining $E_\alpha{}^i$ variables,
except the one for $E_3\!^3$, decouple, leaving a \emph{reduced coupled
system of PDEs} in $x^0$ and $x^3$ for the state vector $E_3\!^3 \oplus
H^{-1} \oplus \bm{S}$, for details, see~\cite{heietal09}.

%Being the simplest inhomogeneous models with oscillatory singularities, the
%$G_2$ models form the natural testing ground for BKL locality, and its
%breaking. Furthermore, as will be discussed below, the $G_2$ models also
%occur naturally in another context, namely solution generating techniques
%that lead to a hierarchically connected set of building blocks for
%understanding and describing oscillatory singularities.

%-----------------------------------------------------------------
\subsubsection*{Invariant subsets induced by conditions on spatial Killing vectors}\label{Subsec:cond}
%-----------------------------------------------------------------

For models with spacelike Killing vectors, further specialized invariant
subset structures can be obtained by imposing conditions on the geometrical
properties of the Killing vector fields. Notable conditions are demands such
as \emph{hypersurface orthogonal} individual Killing vectors, and, in the
$G_2$ case, requiring that the 2-spaces orthogonal to the orbits of the $G_2$
symmetry group are surface forming. This latter case yields the important
invariant \emph{orthogonally transitive} (OT) subset, for which $E_3\!^1 =
E_3\!^2 =0$ and $R_1=0=\Sigma_{23}$. The OT case turns out to be a key subset
for the building blocks needed to describe oscillatory singularities. In
addition special subsets are obtained by taking intersections of the above
subsets and by introducing additional isotropies. Conditions on Killing
vector fields is, however, not the only way of obtaining invariant subsets
from symmetries, invariant subsets also arise if one imposes discrete
isometries. Taken together, symmetry conditions lead to hierarchies of
invariant subsets with different state space dimensions,
see~\cite{waiell97,lim04} for examples.

%-----------------------------------------------------------------
\subsection{Symmetries and invariant boundary correspondence hierarchies}\label{Subsec:symbound}
%-----------------------------------------------------------------

%-----------------------------------------------------------------
\subsubsection*{The local boundary}\label{Subsec:locbound}
%-----------------------------------------------------------------

The system of PDEs given by~\eqref{alleveq} and~\eqref{allconeq} admits an
unphysical invariant boundary subset given by setting $E_\alpha{}^i =0$ (and
hence also $\parb_\alpha =0$), which corresponds to setting the spatial
covariant metric $G^{ij}$ identically to zero. This invariant boundary subset
was previously called the silent boundary (due to that particle horizons form
and shrink to zero size along a timeline towards a spacelike part of a
singularity if ${\cal N}E_\alpha{}^i \rightarrow0$ sufficiently fast), but
since there now is some evidence for that asymptotic silence is associated
with several phenomena, this subset is now called \emph{the local boundary}.
It follows immediately that the equations for the remaining state space
variables $H^{-1}\oplus\bm{S}$ are identical to those in the SH case, and
that the equation for $H^{-1}$ decouple, leaving a coupled system for the
dimensionless Hubble-normalized commutator variables $\bm{S}$ that form the
\emph{dimensionless local boundary state space}. Once the equations are
solved on this reduced state space, they yield a quadrature for $H^{-1}$, but
note that $H^{-1}\rightarrow 0$, where $H^{-1}=0$ is an invariant boundary
subset of the state space $H^{-1}\oplus\bm{S}$, towards the past, as in the
SH case. The only difference with the SH case is that the state space now
consists of an infinite set of copies, one for each spatial point. As a
consequence constants of the motion in the SH case are now replaced with
spatially dependent functions.

%-----------------------------------------------------------------
\subsubsection*{Partially local boundaries}\label{Subsec:partlocbound}
%-----------------------------------------------------------------

The system of PDEs given by~\eqref{alleveq} and~\eqref{allconeq} admits an
unphysical invariant boundary subset given by setting all $E_\alpha{}^i$ to
zero except $E_3\!^3$, and hence also $\parb_1 = \parb_2 = 0$, which
corresponds to setting all components of the spatial covariant metric
$G^{ij}$ except $G^{33}$ identically to zero. It follows immediately that the
equations for the remaining state space $E_3\!^3\oplus H^{-1}\oplus\bm{S}$
are identical to those of the $G_2$ case. We will refer to this subset as
\emph{the partially local $G_2$ boundary}. The difference with the $G_2$ case
is that functions of integration are no longer just functions of $x^3$ but
also of $x^1$ and $x^2$. The situation is therefore completely analogous to
the SH case and the local boundary. A similar statement also holds for models
with one spacelike Killing vector, the so-called $G_1$ models, which hence
give rise to \emph{the partially local $G_1$ boundary}.

As a consequence of the identification of the reduced equation systems for
models with isometry groups and those on corresponding boundary subsets, it
follows that the hierarchy of state spaces induced by imposing further
conditions on isometries, such as the OT $G_2$ case, and by imposing discrete
isometries, translates to the boundary subset state spaces. The boundary
subsets turn out to be essential for past asymptotic dynamics, and hence the
conformally Hubble-normalized state space picture provides a new context for
models with spacelike isometry groups.

%-----------------------------------------------------------------
\subsection{The Lie contraction hierarchy on the local boundary}\label{Subsec:Lielocbound}
%-----------------------------------------------------------------

To simplify the discussion about the local boundary, we use the
identification with the SH equations and therefore talk about SH models
instead. BKL is associated with generic structures, and hence it is the most
general SH models that are of primary interest. These models are the class B
(defined by $A_\alpha\neq 0$) general Bianchi type VI$_{-1/9}$ models and the
class A (defined by $A_\alpha=0$) Bianchi type VIII and IX models, where the
two latter, particularly Bianchi type IX, are the models that have attracted
most interest, and for which we have most rigorous results.

The equations on the dimensionless reduced state space $\bm{S}$ form a
hierarchical invariant subset structure based upon the Bianchi type
classification, where each Bianchi type is associated with an invariant
subset on $\bm{S}$. Furthermore, the Bianchi types that can be obtained from
another Bianchi type by means of Lie contractions, i.e., by setting structure
constants to zero, correspond to invariant boundary subsets of the more
general Bianchi type that can be obtained by setting the corresponding
Hubble-normalized spatial commutator variables to zero. To be more specific,
we will first focus on the class A models.

%-----------------------------------------------------------------
\subsubsection*{Class A models}\label{subsubsec:classA}
%-----------------------------------------------------------------

In the case of the class A vacuum Bianchi models the Codazzi
constraint~\eqref{dlcodazzi} takes the form
$\epsilon_{\alpha}\!^\beta\!_\gamma\Sigma_{\beta\delta}N^{\delta\gamma}=0$,
from which it follows that $N^{\alpha\beta}$ and $\Sigma_{\alpha\beta}$ can
be simultaneously diagonalized (neglecting global topological issues),
moreover, it is possible to diagonalize them in a Fermi frame. This gives a
particularly convenient representation of these models, and we hence set
$N^{\alpha\beta} = {\rm diag}(N_1, N_2, N_3)$, $\Sigma_{\alpha\beta} = {\rm
diag}(\Sigma_1, \Sigma_2, \Sigma_3)$ and $R_\alpha=0$. Once results are
obtained in this frame they can be transformed to an arbitrary Fermi frame by
a general constant orthogonal transformation (a temporally constant
transformation in the local boundary case). This leads to the following
dynamical system on $\bm{S}$ \cite{heiugg09a}:
\begin{subequations}\label{IXeq}
\begin{align}
\label{sig}
\parb_0\Sigma_\alpha & =  2(1-\Sigma^2)\Sigma_\alpha + \sfrac{1}{3}\!\left[ N_\alpha(2N_\alpha - N_\beta
- N_\gamma) - (N_\beta - N_\gamma)^2 \right]\!, \\[0.5ex]
\label{n}
\parb_0N_\alpha & =  -2(\Sigma^2 + \Sigma_\alpha)\,N_\alpha
\qquad\qquad\qquad\qquad\quad\,\, \text{(no sum over $\alpha$)},\\
\label{gauss}
1 &= \Sigma^2 +
\sfrac{1}{12} \Big[ N_1^2 + N_2^2 + N_ 3^2 -
2 \left( N_1N_2 + N_2 N_3 + N_3N_1 \right)\Big],
\end{align}
\end{subequations}
where $(\alpha\beta\gamma) \in \left\{(123),(231),(312)\right \}$
in~\eqref{sig}, and where $\Sigma_1 + \Sigma_2 + \Sigma_3 = 0$, $\Sigma^2 =
\frac16(\Sigma_1^2 + \Sigma_2^2 + \Sigma_3^2)$.

The different class A models are characterized as follows: Bianchi type IX
(VIII) has $N_1N_2N_3>0$ ($N_1N_2N_3<0$) and a reduced dimensionless
4-dimensional state space; Bianchi type VII$_0$ (VI$_0$) $N_\alpha N_\beta
>0$ ($N_\alpha N_\beta <0$), $N_\gamma = 0$ ($(\alpha\beta\gamma) = (123)$,
and cycle), with a 3-dimensional state space; Bianchi type II
$N_\alpha=N_\beta=0$, $N_\gamma \neq 0$ ($(\alpha\beta\gamma) = (123)$, and
cycle), leading to a 2-dimensional state space, and finally Bianchi type I is
characterized by $N_1=N_2=N_3=0$, which leads to the 1-dimensional so-called
Kasner circle $\mathrm{K}^{\ocircle}$ of fixed points, discussed below. Hence
a Bianchi type IX (VIII) model has three Bianchi type VII$_0$ (two VI$_0$ and
one VII$_0$) invariant boundary subsets, while each of those subsets have two
Bianchi type II invariant boundary subsets, and, finally, each Bianchi type
II subset in this hierarchy of subsets has the Bianchi type I Kasner circle
$\mathrm{K}^{\ocircle}$ as an invariant boundary subset.

Setting a spatial commutator variable $N_\alpha$ to zero corresponds to
setting an associated structure constant to zero, which leads to a lower
Bianchi type via a so-called Lie contraction~\cite{jan01}. Thus the above
dynamical system exhibits a `\emph{Lie contraction boundary subset
hierarchy.}' The hierarchical Lie contraction boundary subset structure is
central for the asymptotic dynamics towards the initial singularity. Setting
a variable $N_\alpha$ to zero is associated with increasing the dimension of
the automorphism group with one.\footnote{In the present context automorphism
transformations are the linear constant transformations of a symmetry adapted
spatial frame that leave the structure constants unchanged.} The
\emph{kinematical} consequence of this is that a given Lie contracted
boundary subset of~\eqref{IXeq} describes the true degrees of freedom of the
associated Bianchi type, where integrating $\parb_{0}H^{-1} = (1 +
2\Sigma^2)H^{-1}$, once a solution of~\eqref{IXeq} has been obtained, yields
the scale parameter of the solution. Due to this, the metric can be
algebraically constructed from the solution on $H^{-1}\oplus\bm{S}$ by means
of group theoretical methods, see~\cite{janugg99}.

More importantly, however, are the \emph{dynamical} implications of the group
of automorphisms and scale transformations. As explicitly shown
in~\cite{heiugg10}, on each level in the Lie contraction boundary subset
hierarchy the combined scale-automorphism group induces monotone functions,
and even constants of the motion at the bottom of the hierarchy. The
resulting \emph{hierarchy of monotone functions} pushes the dynamics towards
the past singularity to boundaries of boundaries in the hierarchy, where the
solutions at the bottom of the hierarchy, i.e., those of Bianchi types I and
II, are completely determined by the symmetries induced by the
scale-automorphism groups of these models. The dynamical evolution towards
the initial singularity is therefore to a large extent governed by structures
induced by the scale-automorphism groups on the different levels in the Lie
contraction boundary subset hierarchy. Since the automorphisms in the present
SH context correspond to the spatial diffeomorphism freedom that respects the
symmetries of the various Bianchi models, it therefore follows that \emph{the
dynamical evolution towards the initial singularity is partly determined by
physical first principles, namely scale-invariance and general covariance}.

In the case of matter sources, hierarchies become even more important than in
the vacuum case. Then, in addition to Lie contractions, one also have source
contractions, where vacuum is at the bottom of the source hierarchy. For each
level of the source contraction hierarchy the scale-automorphism group yield
different structures, such as monotone functions, leading to restrictions on
asymptotic dynamics; see~\cite{heiugg10} where this general feature is
exemplified explicitly for an orthogonal perfect fluid with a linear equation
of state in the case of diagonal class A Bianchi models.

%-----------------------------------------------------------------
\subsubsection*{Class B models: The general Bianchi type VI$_{-1/9}$ case}\label{subsubsec:classB}
%-----------------------------------------------------------------

The class B Bianchi models satisfy
\begin{equation}
A_{\beta} N_{\alpha}{}^{\beta} = 0,\qquad A_{\alpha}A^\alpha =\sfrac{1}{2} h \left((N^\alpha\!_\alpha)^2 -
N^\alpha\!_\beta\,N^\beta\!_\alpha\right),
\end{equation}
where $h$ is a constant parameter that characterizes the Bianchi type VI$_h$
and VII$_h$ models, which follows from the analogous structure constant
relation. But $h$ can also be viewed as a constant of the motion on the state
space $\bm{S}$, and hence $h$ is a spatial function on the class B local
boundary. In class B the vacuum Codazzi constraint~\eqref{dlcodazzi} gives
$3A_{\beta}\Sigma_{\alpha}{}^{\beta} =
\epsilon_{\alpha}{}^{\beta\gamma}\Sigma_{\beta}{}^{\delta}N_{\delta\gamma}$.
For all class B models, except type V and some special type VI$_h$ models
(including Bianchi type III), $3A_{\beta}\Sigma_{\alpha}{}^{\beta} =
\epsilon_{\alpha}{}^{\beta\gamma}\Sigma_{\beta}{}^{\delta}N_{\delta\gamma}\neq
0$, and as a consequence, in contrast to class A, $\Sigma_{\alpha\beta}$ and
$N_{\alpha\beta}$ are not in general simultaneously diagonalizable.
Kinematically the Bianchi type VI$_h$ and VII$_h$ models are the most general
class B models, but for $h=-1/9$ the Codazzi constraints become degenerate,
which permits $\Sigma_{\alpha\beta}$ to have one more independent component
than for the other type VI$_h$ and VII$_h$ models, making the state space
4-dimensional, i.e., type VI$_{-1/9}$ is as general as Bianchi types VIII and
IX~\cite{ellmac69,waiell97,hewetal03}. Moreover, the extra degree of freedom
these models admit leads to an oscillatory singularity, in contrast to all
other class B vacuum models, which are past asymptotically self-similar. As
regards class B, it is thus the general Bianchi type VI$_{-1/9}$ vacuum
models that are of interest for oscillatory singularities, furthermore, they
exhibit some other oscillatory features than Bianchi types VIII and IX,
making them interesting as toy models for general oscillatory singularities.

The reduced dimensionless dynamical system for the general Bianchi type
VI$_{-1/9}$ vacuum models (and its local boundary analogue) can be written as
follows (choosing an Iwasawa frame so that $(R_1,R_2,R_3) = -(\Sigma_{23},0,
\Sigma_{12})$ and so that the remaining non-zero variables are $\Sigma_\alpha
= \Sigma_{\alpha\alpha}$, $A_3 = A$, $N_{11}=N_1$, $N_{12}= 3A$,
see~\cite{heietal09}):

\textit{Evolution equations}:
\begin{subequations}\label{VIevoleq}
\begin{align}
\parb_0\Sigma_1 &= - 2\Omega_\mathrm{k}\Sigma_1 + 2R_3^2 - \sfrac23 N_1^2, \label{S1VI}\\
\parb_0\Sigma_2 &= - 2\Omega_\mathrm{k}\Sigma_2 - 2(R_3^2 - R_1^2) + \sfrac13 N_1^2 - 12A^2, \label{S2VI}\\
\parb_0\Sigma_3 &= - 2\Omega_\mathrm{k}\Sigma_3 - 2R_1^2 + \sfrac13 N_1^2 + 12A^2, \label{S3VI}\\
\parb_0R_1 &= (- 2\Omega_\mathrm{k} + \Sigma_3 - \Sigma_2)R_1, \label{R1VI}\\
\parb_0R_3 &= (- 2\Omega_\mathrm{k} + \Sigma_2 - \Sigma_1)R_3 + 4N_1A, \label{R3VI}\\
\parb_0N_1 &= 2(\Sigma^2 + \Sigma_1)N_1 - 12R_3A, \label{N1VI}\\
\parb_0A &= (2\Sigma^2 - \Sigma_3)A. \label{AVI}
\end{align}
\end{subequations}

\textit{Constraint equations}:
\begin{subequations}\label{VIconstreq}
\begin{align}
0 &= 1 - \Sigma^2 - \Omega_\mathrm{k}, \label{GaussVI}\\
0 &=  6\Sigma_1A + R_3N_1,\label{CodazziVI}\\
0 &= \Sigma_1 + \Sigma_2 + \Sigma_3, \label{ShearconstrVI}
\end{align}
\end{subequations}
where $\Sigma^2 = \frac16(\Sigma_1^2 + \Sigma_2^2 + \Sigma_3^2 + 2R_1^2 +
2R_3^2)$ and $\Omega_\mathrm{k} = \sfrac{1}{12}N_1^2 + 4A^2$.

To obtain the dynamical hierarchical structure induced by the
scale-automorphism group hierarchy in class A, Hamiltonian techniques were
used in~\cite{heiugg10}, but this was only for practical reasons, the results
do not depend on them since they are consequences of the scale-automorphism
group hierarchy. The class B models do not, in general, have a Hamiltonian
formulation (without extra non-potential forces). For this reason an
exploration of the dynamical consequences of the scale-automorphism group
would have to take a more direct approach. Moreover, due to that the models
are non-diagonal, especially the general Bianchi type VI$_{-1/9}$ models,
these models, in contrast to the class A vacuum models, also involve
off-diagonal automorphisms. A natural step in a systematic analysis of the
class B models, and especially the general Bianchi type VI$_{-1/9}$ models,
would therefore be an analysis of the dynamical implications of the
scale-automorphism group, a task that remains to be done.

%%%%%%%%%%%%%%%%%%%%%%%%%%%%%%%%%%%%%%%%%%%%%%%%%%%%%%%%%%%%%%%%%%
\section{BKL, transitions, concatenation, chains and discretized maps}\label{Sec:concat}
%%%%%%%%%%%%%%%%%%%%%%%%%%%%%%%%%%%%%%%%%%%%%%%%%%%%%%%%%%%%%%%%%%

%-----------------------------------------------------------------
\subsection{The generalized Kasner solution}\label{Subsec:genKasner}
%-----------------------------------------------------------------

The BKL results in~\cite{bkl82} were obtained by means of a synchronous
coordinate frame, and using the coordinate components of the metric and its
synchronous time derivative as basic variables as starting point. However,
BKL also emphasizes the role of spatial scale factors, obtained by
diagonalizing the spatial frame. As discussed below, such diagonalizations
can be accomplished in many different ways, which lead to different
descriptions of asymptotic dynamics, although, presumably, the \emph{local}
gauge-invariant features are the same. In all cases, the BKL analysis starts
with inserting the generalized Kasner solutions (obtained by dropping all
spatial derivatives in a diagonalized spatial frame and solving for the scale
factors) into the different terms in the full Einstein field equations in the
chosen spatial frame. It is then found that certain terms grow and eventually
become non-negligible `triggering transitions' (in the present notation),
leading to Kasner oscillations.

The BKL picture thus relies on the central importance of the Kasner
solutions. In the SH case these are the Bianchi type I solutions which are
obtained by setting $N^{\alpha\beta}=0$ and $A_\alpha=0$. Doing this, in the
SH case as well as on the local boundary, leads to that~\eqref{dlsigdot}
takes the form
\begin{equation}\label{shearBI}
\parb_{0}\Sigma_{\alpha\beta} =
2\epsilon^{\gamma\delta}{}_{\langle\alpha}\,\Sigma_{\beta\rangle\gamma}\,R_{\delta},
\end{equation}
since the Gauss constraint~\eqref{dlgauss} gives $\Sigma^2 = 1$ and hence $q=
2\Sigma^2 = 2$.

In the present framework the metric scale factors of BKL are associated with
temporal integration of the diagonal components of the conformally
Hubble-normalized expansion tensor $\Theta_{\alpha\beta}$, which can be
expressed in terms of the deceleration parameter $q$ and the
Hubble-normalized shear $\Sigma_{\alpha\beta}$ as follows:
\begin{equation}
\Theta_{\alpha\beta} =
-q\delta_{\alpha\beta} + \Sigma_{\alpha\beta}.
\end{equation}
Furthermore, from the above equation it follows that if one focusses on
evolution that quotes out overall evolution as regards scale factors, as done
in `asymptotic billiard approaches'~\cite{chi72,grav73,dametal03}, then this
is associated with the diagonal shear variables
$(\Sigma_{1},\Sigma_{2},\Sigma_{3}):=(\Sigma_{11},\Sigma_{22},\Sigma_{33})$.

In all `metric scale factor' approaches (as explicitly illustrated by spatial
Fermi and Iwasawa frames below), the equation~\eqref{shearBI} admits a
special solution characterized by $R_1=R_2=R_3=0$, for which furthermore
$\Sigma_{\alpha\beta}$ is diagonal, i.e., they all admit a special case for
which the spatial frame is a Fermi frame in which the Hubble-normalized shear
is diagonalized. As a consequence $\parb_{0}\Sigma_{\alpha} = 0$, and hence
$\Sigma_{\alpha} = \hat{\Sigma}_{\alpha}$, where $\hat{\Sigma}_{\alpha}$ are
constants in the SH case and temporal constants on the local boundary, i.e.
spatial functions. This feature is equivalent to the statement that the
Kasner solutions appear as fixed points on the dimensionless state space
$\bm{S}$ (fixed points are also often referred to as equilibrium points,
places where the flow of a dynamical system is zero). Since the Kasner
solutions are fixed points on the scale-invariantly dimensionless state space
$\bm{S}$, it follows that they are scale-invariant and admit a spacetime
transitive homothety group, as was formally shown
in~\cite{rosjan85,janros86,hsuwai86}. This is one of the advantages of the
Hubble-normalized state space approach to SH cosmology. However, this feature
becomes even more pertinent in the conformally Hubble-normalized approach to
general inhomogeneous models, since the SH/local boundary correspondence
leads to a correspondence between models with spacetime transitive
self-similar symmetry groups and fixed points on the local boundary. Special
models that asymptotically approach such a self-similar model in the SH case
are said to be \emph{asymptotically self-similar}, leading to that special
inhomogeneous models that approach fixed points on the local boundary are
naturally characterized as being \emph{locally asymptotically self-similar}
(in the Kasner case they are often called, less geometrically, asymptotically
velocity dominated).

The $\hat{\Sigma}_{\alpha}$ values can be expressed as follows:
\begin{equation}
\hat{\Sigma}_{\alpha} =: 3p_\alpha - 1,
\end{equation}
where $p_\alpha$ are the Kasner parameters in the SH case and the so-called
generalized Kasner parameters in the local boundary case. These parameters,
which satisfy $p_1 + p_2 + p_3=1 = p_1^2 + p_2^2 + p_3^2$, as a consequence
of that $\Sigma_{\alpha\beta}$ is trace-free and that $\Sigma^2
=\frac16(\hat{\Sigma}_1^2 + \hat{\Sigma}_2^2 + \hat{\Sigma}_3^2) =1$,
describe the \emph{Kasner circle} $\mathrm{K}^{\ocircle}$ on $\bm{S}$. It
follows that a shear diagonalized Fermi frame yields
\begin{equation}\label{genKasnereq}
\parb_0 H^{-1} = 3H^{-1},\quad \parb_0 E_\alpha{}^i = 3(1-p_\alpha)E_\alpha{}^i \quad \Rightarrow \quad
\parb_0 e^\alpha\!_i = 3p_\alpha\, e^\alpha\!_i \quad (\text{no sum over $\alpha$}),
\end{equation}
where these equations are obtained by inserting the `diagonalized' Kasner
solution on the reduced state space $\bm{S}$ into the equation~\eqref{Heq}
for $H^{-1}$ and into that of~\eqref{dl13comts} for $E_\alpha{}^i$, which can
be viewed as a perturbation of the local boundary into the physical state
space. Choosing a synchronous time variable $t$ defined by $N=1$ gives ${\cal
N} = H$ and $\parb_0 = H^{-1}\partial_t$, which leads to $\partial_t H^{-1} =
3$ and hence $H^{-1} = 3t + C$, where $C$ is a constant in the SH case, while
it is a spatial function in the local boundary case. Using the gauge freedom
to choose a foliation such that the singularity occurs simultaneously at
$t=0$ gives $H^{-1} = 3t$. This leads to $t\partial_t e^\alpha\!_i = p_\alpha
e^\alpha\!_i$, which gives $e^\alpha\!_i = \hat{e}^\alpha\!_i t^{p_\alpha}$,
where $\hat{e}^\alpha\!_i$ are arbitrary spatial functions in the perturbed
local boundary case. This results in the line element
\begin{equation}\label{genKasnerlineelement}
ds^2 = - dt^2 + \sum_\alpha t^{2p_\alpha}(\hat{e}^\alpha\!_i dx^i)(\hat{e}^\alpha\!_j dx^j),
\end{equation}
which gives the BKL result of the generalized Kasner metric where
$\hat{e}^\alpha\!_i$ can be identified with the `BKL Kasner axes.'

There is another way this result can be derived. Consider a given timeline
and choose a time variable $\tau$ along this timeline so that ${\cal N}=1$,
i.e., $N = H^{-1}$, and $\parb_0 = \partial_\tau$. Then~\eqref{genKasnereq}
yields $\partial_\tau H^{-1} = 3H^{-1}$ and $\partial_\tau e^\alpha\!_i =
3p_\alpha\, e^\alpha\!_i$, which results in $H^{-1} =
\hat{H}^{-1}\exp(3\tau)$, $e^\alpha\!_i =
\tilde{e}^\alpha\!_i\exp(3p_\alpha\tau)$ where $\tilde{e}^\alpha\!_i$ are
temporal constants. It follows that the singularity occurs at $\tau
\rightarrow - \infty$ and that $t - \hat{t} = \hat{H}^{-1}\exp(3\tau)/3$. By
choosing a gauge in which the temporal constant $\hat{t}$ is set to zero, so
that the singularity occurs at $t=0$, and by scaling $\tilde{e}^\alpha\!_i$
with the temporal constant $\hat{H}^{-1}$ appropriately, we obtain the
previous result~\eqref{genKasnerlineelement}.

Due to axis permutations, the Kasner circle $\mathrm{K}^{\ocircle}$ is
naturally divided into six equivalent sectors, denoted by permutations of the
triple $(123)$ where sector $(\alpha\beta\gamma)$ is defined by $p_\alpha
\in(-\frac13, 0) < p_\beta \in(0, \frac23) < p_\gamma \in(\frac23, 1)$. The
boundaries of the sectors are six special points that in the SH case exhibit
multiply transitive symmetry groups and belong to the class of locally
rotationally symmetric (LRS) solutions (they are even plane symmetric, and
hence also axially symmetric, a property that will feature in the discussion
in Section~\ref{Sec:disc}),
\begin{subequations}
\begin{alignat}{2}
\mathrm{Q}_\alpha: \quad & (p_\alpha,p_\beta,p_\gamma)\,
&=&\, (-\textfrac{1}{3},\textfrac{2}{3},\textfrac{2}{3}),\\
\mathrm{T}_\gamma: \quad & (p_\alpha,p_\beta,p_\gamma)\, &=&\, (0,0,1).
\end{alignat}
\end{subequations}
The $\mathrm{Q}_\alpha$, $\alpha = 1,2,3$, points yield three equivalent LRS
solutions with non-flat geometry, while the \textit{Taub points}
$\mathrm{T}_\gamma$, $\gamma = 1,2,3$, correspond to the Taub representation
of Minkowski spacetime.

Finally, note that $\parb_0 E_\alpha{}^i = 3(1-p_\alpha)E_\alpha{}^i$ leads
to that $E_\alpha{}^i \rightarrow 0$ towards the past, except at the Taub
points, which suggests that if solutions are `dominated' by non-Taub Kasner
states they might approach the local boundary, \emph{if} $E_\alpha{}^i
\rightarrow 0$ \emph{faster} than spatial derivatives might grow, which turns
out \emph{not} to be the case for the recurring spikes discussed later.

To obtain further results require an explicit choice of spatial frame. We
begin with a Fermi frame.

%-----------------------------------------------------------------
\subsection{Transitions, concatenation and chains in a Fermi frame}\label{Subsec:concat}
%-----------------------------------------------------------------
%Transitions, concatenation, chains and oscillations on the local boundary

In this subsection we pursue outlining how the BKL results can be obtained in
the conformally Hubble-normalized state space picture in a Fermi frame.
Choosing a Fermi frame, i.e., $R_\alpha =0$, leads to that~\eqref{shearBI}
results in $\parb_{0}\Sigma_{\alpha\beta} = 0$, and hence that
$\Sigma_{\alpha\beta} = \hat{\Sigma}_{\alpha\beta}$, where
$\hat{\Sigma}_{\alpha\beta}$ are constants in the SH case and temporal
constants on the local boundary, i.e. spatial functions. The symmetric matrix
$\hat{\Sigma}_{\alpha\beta}$ can therefore be diagonalized by making an
appropriate (temporally) constant orthogonal transformation of the spatial
frame, which results in the eigenvalues $\hat{\Sigma}_{\alpha} =: 3p_\alpha -
1$, which leads to the results derived above.

The instability result of BKL of the generalized Kasner solution follows from
perturbing $\mathrm{K}^{\ocircle}$ \emph{on} the dimensionless state space
$\bm{S}$ on the local boundary; a linearization in a shear diagonalized Fermi
frame yields
\begin{subequations}\label{geomstabeq}
\begin{alignat}{2}
\parb_0A_\gamma &= 3(1-p_\gamma)A_\gamma,&
\label{Astab}\\
\parb_0N_{\alpha} &= 6p_\alpha N_{\alpha}
&\quad \text{where} &\qquad\, N_\alpha:=N_{\alpha\alpha},
\label{Ndiastab}\\
\parb_0N_{\alpha\beta} &=
3(1- p_\gamma)N_{\alpha\beta} &\quad \text{where} &\qquad (\alpha\beta\gamma)
= (123)\,\, \text{and cycle}\label{Noffdstab}
\end{alignat}
\end{subequations}
(there also are zero eigenvalues due to that $\parb_{0}\Sigma_{\alpha\beta} =
0$, which correspond to a center subspace). All the eigenvalues associated
with~\eqref{geomstabeq} are strictly positive, and thus stable \emph{towards
the past}, except at (i) the Taub points, which reflects that they are not
transversally hyperbolic fixed points of the dynamical system on the local
boundary, and (ii) at the arc of the Kasner circle, denoted by
$\mathrm{K}^{\ocircle}_\alpha$, determined by $p_\alpha <0$ in
eq.~\eqref{Ndiastab}, which corresponds to the union of sectors
$(\alpha\beta\gamma)$ and $(\alpha\gamma\beta)$ and the point
$\mathrm{Q}_\alpha$, which leads to that $N_\alpha$ is unstable towards the
past; hence $\mathrm{K}^{\ocircle}$ consists of three equivalent arcs,
$\mathrm{K}^{\ocircle}_1$, $\mathrm{K}^{\ocircle}_2$ and
$\mathrm{K}^{\ocircle}_3$, separated from each other by the Taub points,
where each arc is associated with a towards the past unstable $N_\alpha$
variable, which in turn corresponds to a 1-dimensional past unstable subspace
on $\bm{S}$ at each fixed point on $\mathrm{K}^{\ocircle}_\alpha$. Neglecting
the problems the Taub points pose for the moment, it follows that
asymptotically all conformally Hubble-normalized spatial connection variables
tend to zero at $\mathrm{K}^{\ocircle}$ except for a single diagonal
component of $N^{\alpha\beta}$ at each arc $\mathrm{K}^{\ocircle}_\alpha$, a
situation that corresponds to that in a Fermi frame precisely one of the
eigenvalues of $N^{\alpha\beta}$ is always unstable towards the past.
Following the nomenclature of~\cite{heietal09}, unstable variables towards
the past are referred to as \emph{trigger variables}.\footnote{The BKL
procedure corresponds to inserting the generalized Kasner solution into the
spatial 3-curvature and studying how it destabilizes the generalized Kasner
solutions. In the present approach the analogous leading order expression for
the conformally Hubble-normalized spatial 3-curvature is obtained by
solving~\eqref{geomstabeq}.}

%-----------------------------------------------------------------
\subsubsection*{Bianchi type II transitions on the local boundary}\label{Subsec:transitions}
%-----------------------------------------------------------------

Setting all spatial commutator variables to zero except for a single trigger
variable $N_\alpha$ yields an invariant subset on the local boundary that
corresponds to the Bianchi type II models in the SH case. The equations for
these models, on $\bm{S}$ in a shear diagonalized Fermi frame, are given
by~\eqref{IXeq} by setting $N_\alpha \neq 0$, $N_\beta=N_\gamma =0$, where
$(\alpha\beta\gamma)=(123)$ or a permutation thereof, where we refer to the
associated subset by $\mathcal{B}_{N_\alpha}$. These equations are easily
solved, see e.g.~\cite{heietal09}, and yield a 1-parameter set of
\emph{heteroclinic orbits}, i.e. solution trajectories that connect two
distinct fixed points (in the present case, two different Kasner fixed
points). More precisely, consider an equilateral triangle that circumscribes
$\mathrm{K}^{\ocircle}$ in such a way so that it is tangential to the three
Taub points. Then each Bianchi type II solution, projected onto the
Hubble-normalized shear plane (since $\Sigma_1+\Sigma_2+\Sigma_3 =0$) of
$\mathcal{B}_{N_\alpha}$, is a straight line inside $\mathrm{K}^{\ocircle}$
that can be extended outside $\mathrm{K}^{\ocircle}$ through the corner of
the equilateral triangle closest to $\mathrm{Q}_\alpha$, see
e.g.~\cite{waiell97}.

In~\cite{heietal09} the Bianchi type II heteroclinic orbits were referred to
as $\cT_{N_\alpha}$ (Kasner) \emph{transitions} since their past and future
limits are two distinct points on $\mathrm{K}^{\ocircle}$, thus yielding a
map between different Kasner states. To describe this map it is convenient to
parameterize the (generalized) Kasner parameters $p_\alpha$ on
$\mathrm{K}^{\ocircle}$ in terms of a (generalized) Kasner parameter $u$.
Instead of following BKL and defining it via $p_\alpha \leq p_\beta \leq
p_\gamma$ it can be defined in a gauge-invariant way as follows: Consider the
quantity
\begin{equation}
p_1p_2p_3 = - \frac{u^2(1+u)^2}{(1+u+u^2)^3}.
\end{equation}
Due to that $p_1p_2p_3$ is monotone in $u$ this relation defines $u$
implicitly, and since it can be shown that $p_1p_2p_3$ can be constructed
from the Weyl scalars it follows that $u$ gauge-invariantly describes the
different Kasner states. The parameter $u \in [1,\infty]$, where $u=1$
corresponds to the $\mathrm{Q}_\alpha$ Kasner state while $u=\infty$ yields
the Taub Kasner state.

Towards the past, the transition $\mathcal{T}_{N_\alpha}$ gives rise to a map
between two different Kasner points on $\mathrm{K}^{\ocircle}$ (where the map
is defined to be the identity at $\mathrm{T}_\beta$ and $\mathrm{T}_\gamma$),
see e.g.~\cite{heietal09,ugg13a}; expressing the result in the
gauge-invariant Kasner parameter $u$ yields the BKL Kasner
map~\cite{khaetal85,heietal09}:
\begin{equation}\label{BKLMap}
u_+  \:= \:
\left\{\begin{array}{ll}
u_- - 1 & \qquad \text{if}\quad u_- \geq 2 \\[1ex]
(u_- - 1)^{-1} &\qquad  \text{if} \quad 1 \leq u_- < 2
\end{array}\right. ,
\end{equation}
where $u=u_-$ and $u_+$ are the initial and final Kasner states,
respectively, in the direction towards the past.

The `generalized' Bianchi type II solution can be obtained by taking the
dimensionless Bianchi type II solution on the local boundary on the state
space $\bm{S}$ and inserting it into the equations for $H^{-1}$ and
$E_\alpha{}^i$, i.e.~\eqref{Heq} and ~\eqref{dl13comts}, respectively, thus
perturbing the local boundary, in a similar way as the generalized Kasner
solution was obtained above. One can continue the perturbative expansion by
linearly perturbing this solution away further from the local boundary, which
yields new ODEs, since partial spatial derivatives act only `passively' on
the spatial functions of integration that are obtained in solving the
equations that arise from the perturbative expansion, order by order. This
gives e.g. the BKL results concerning the rotation and asymptotic freezing of
Kasner axes, which are associated with the following feature. \emph{On} the
local boundary one can choose a frame in the vacuum case so that for all of
class A, including Bianchi types I and II, $N^{\alpha\beta}$ and
$\Sigma_{\alpha\beta}$ both become diagonal, a feature that is due to that
the Codazzi constraint~\eqref{dlcodazzi} in this case is given by $
\epsilon_{\alpha}{}^{\beta\gamma}\Sigma_{\beta}{}^{\delta}N_{\delta\gamma}
=0$. Perturbations of the Bianchi type II solutions on the local boundary
into the physical state space give rise to non-zero Hubble-normalized spatial
frame derivatives, resulting in $
\epsilon_{\alpha}{}^{\beta\gamma}\Sigma_{\beta}{}^{\delta}N_{\delta\gamma}
\neq 0$, which prevent simultaneous diagonalization. BKL assumes that
solutions can be described as increasingly small perturbations of sequences
of Bianchi type II solutions on the local boundary, which leads to $
\epsilon_{\alpha}{}^{\beta\gamma}\Sigma_{\beta}{}^{\delta}N_{\delta\gamma}
\rightarrow 0$, which hence explains the BKL asymptotic freezing effect of
the Kasner axes. A similar statement holds for the electric and magnetic Weyl
tensors, which offers a more geometric way of stating the rotation and
asymptotic freezing of Kasner axes properties.

Next, we describe the recent progress that has been achieved as regards SH
dynamics (and hence also for dynamics \emph{on} the local boundary) for
Bianchi types VIII and IX in a $\Sigma_{\alpha\beta}$ and $N^{\alpha\beta}$
diagonalized Fermi frame.

%-----------------------------------------------------------------
\subsubsection*{Global past dynamics results for Bianchi types VIII and IX}\label{Subsec:globe}
%-----------------------------------------------------------------

Although the Lie contraction hierarchy limits asymptotic dynamics, it does
not uniquely determine it, thus making the endeavor of producing theorems
about generic initial singularities in SH models a non-trivial task. The
first theorems in this area concerning oscillatory singularities, based on
earlier work in~\cite{waiell97} and by results obtained by
Rendall~\cite{ren97}, were produced by Ringstr\"om in 2000~\cite{rin00} and
2001~\cite{rin01}. In the first of these papers it was shown that a generic
Bianchi type VIII or IX solution cannot converge to a Taub point on
$\mathrm{K}^\ocircle$ and that the past limit set contains at least two
distinct points on $\mathrm{K}^\ocircle$, of which at least one is not a Taub
point. As a consequence the past singularity in these models must be
oscillatory and a scalar curvature singularity. In the second paper it was
proved that the past attractor for the vacuum Bianchi type IX models,
$\mathcal{A}_{\mathrm{IX}}$, \emph{resides} on the union of
$\mathrm{K}^\ocircle$ and the Bianchi type II subsets, i.e.,
\begin{equation}\label{AIXtheo}
\mathcal{A}_{\mathrm{IX}} \in \overline{\bf B}_\mathrm{II} := \mathrm{K}^\ocircle \cup
\mathcal{B}_{N_1} \cup
\mathcal{B}_{N_2} \cup
\mathcal{B}_{N_3},
\end{equation}
which, alternatively, can be expressed as that a generic solution satisfies
\begin{equation}\label{rinthmeq}
N_1 N_2 + N_2 N_3 + N_3 N_1 \:\rightarrow \:0
\end{equation}
towards the initial singularity in Bianchi type IX. Furthermore, it was shown
in~\cite{rin01} that $\mathcal{A}_{\mathrm{IX}}$ contains at least three
distinct non-Taub points on $\mathrm{K}^\ocircle$. It is noteworthy that
there exists, so far, no such theorem for Bianchi type VIII, especially since
\emph{both BKL and the Hamiltonian billiard approach}, used in
e.g.~\cite{dametal03}, \emph{assume that~\eqref{rinthmeq} holds generically}.
This is clearly a highly non-trivial, although plausible, assumption, which
presumably requires elaboration on how to measure `generic.' Moreover, as
discussed in~\cite{heiugg09a}, the above `attractor theorem' says nothing
about \emph{how} a generic solution asymptotically approaches
$\mathcal{A}_{\mathrm{IX}}$, nor if all of $\mathcal{A}_{\mathrm{IX}}$ is
really the past attractor. The results of Ringstr\"om therefore say e.g.
nothing about if the map~\eqref{BKLMap} has any relevance for generic
singularities. The proof of Ringstr\"om's attractor theorem does not fully
use the structure of the Lie contraction hierarchy, a shorter proof, making
more use of these structures as well as using different bounded variables, is
given in~\cite{heiugg09b}. To proceed further, however, it is necessary to
take a closer look at the structures on $\overline{\bf B}_\mathrm{II}$, which
leads to the concepts of concatenation and heteroclinic chains.

%-----------------------------------------------------------------
\subsubsection*{Concatenation and heteroclinic chains on $\overline{\bf B}_\mathrm{II}$}\label{Subsec:conchetero}
%-----------------------------------------------------------------

In the diagonalized Bianchi types VIII and IX cases the transitions
$\mathcal{T}_{N_1}$, $\mathcal{T}_{N_2}$ and $\mathcal{T}_{N_3}$ on $\bm{S}$
can be uniquely \emph{concatenated} on $\overline{\bf B}_\mathrm{II}$ towards
the past by identifying the `final' fixed point of one transition with the
`initial' fixed point of another transition. Concatenating a sequence of such
orbits towards the past, obtained by means of the above described equilateral
triangle circumscribing $\mathrm{K}^{\ocircle}$, see
e.g.~\cite{waiell97,heiugg09a}, yields a \emph{heteroclinic chain}, which, in
general, is infinite. We refer to the heteroclinic chains in the diagonalized
class A case on $\overline{\bf B}_\mathrm{II}$ as \emph{Mixmaster chains}.

Note that heteroclinic chains, obtained by joining solutions by means of
their asymptotics into chains of solutions connected via fixed points, are
not solutions to the Einstein equations themselves. Instead they are the
rigorous dynamical systems formulation of the heuristic BKL concept of
piecewise joined solutions. Note also that since the local boundary is
effectively a finite dimensional system, although it really is an infinite
set of copies of the same dynamical system, one copy for each spatial point,
this concept is still valid on this boundary subset. However, a solution on
the local boundary is in general described by several heteroclinic chains,
one for each spatial point.

%-----------------------------------------------------------------
\subsection{Heteroclinic chain discretization: Kasner maps}\label{Subsec:dicrete}
%-----------------------------------------------------------------

The Mixmaster chains induce iterations of the gauge-invariant
map~\eqref{BKLMap}.\footnote{The heteroclinic Mixmaster chains also induce
iterations of the map that takes one Kasner point to another on
$\mathrm{K}^{\ocircle}$, for which the permutation freedom has not been
quoted out; for an analytic description, see~\cite{ugg13a}.} Let
$l=0,1,2,\ldots$ and let $u_l$ denote the initial Kasner state of the
$l$\raisebox{0.7ex}{\small th} transition (time direction towards the past),
then the iterated BKL Kasner map is given by:
\begin{equation}\label{Kasnermap}
u_l \:\,\xrightarrow{\;\text{$l$\raisebox{0.5ex}{th} transition}\;}\:\, u_{l+1}:
\qquad\quad
u_{l+1} \:= \: \left\{\begin{array}{ll}
u_l - 1 & \qquad \text{if}\quad u_l \in[2,\infty), \\[1ex]
(u_l - 1)^{-1} &\qquad  \text{if} \quad u_l \in [1,2].
\end{array}\right.
\end{equation}
%
%Since each value of the Kasner parameter $u \in(1,\infty)$ represents an
%equivalence class of six Kasner fixed points, the Kasner map can be regarded
%as the map induced by the Mixmaster map on these equivalence classes via the
%equivalence relation.
In a sequence $(u_l)_{l=0,1,2,\ldots}$ that is generated
by~\eqref{Kasnermap}, each Kasner state $u_l$ is called a \textit{Kasner
epoch}. Every sequence $(u_l)_{l=0,1,2,\ldots}$ possesses a natural partition
into pieces called \emph{Kasner era}s with a finite number of epochs. An era
consists of a sequence of monotonically decreasing values of $u$ that begins
with a maximal value $u_{\lin}$, generated from $u_{\lin-1}$ by $u_{\lin} =
(u_{\lin-1}-1)^{-1}$, and continues with a sequence of Kasner parameters
obtained via $u_l \mapsto u_{l+1} = u_l - 1$; it ends with a minimal value
$u_{\lout}$ that satisfies $1 < u_{\lout} < 2$, so that $u_{\lout+1} =
(u_{\lout}-1)^{-1}$ begins a new era~\cite{bkl70}, as exemplified by
\begin{equation}\label{phases}
\underbrace{3.41 \rightarrow 2.41 \rightarrow 1.41}_{\text{\scriptsize era}}  \rightarrow
\underbrace{2.44 \rightarrow 1.44}_{\text{\scriptsize era}}  \rightarrow
\underbrace{2.27 \rightarrow 1.27}_{\text{\scriptsize era}}  \rightarrow \ldots
\end{equation}

Denoting the initial and maximal value of the Kasner parameter $u$ in era
number $s$ (where $s = 0,1,2,\ldots$) by $\u_s$,  and decomposing $\u_s$ into
its integer $k_s = [\u_s]$ and fractional $x_s = \{\u_s\}$ parts,
gives~\cite{bkl70,khaetal85}
\begin{equation}\label{usdecomp}
\u_s = k_s + x_s,
\end{equation}
where $k_s$ represents the discrete length and number of Kasner epochs of era
$s$. The final and minimal value of the Kasner parameter in era $s$ is given
by $1 + x_s$, which implies that era number $(s+1)$ begins with
\begin{equation}\label{eramap}
\u_{s+1} = \frac{1}{x_s} = \frac{1}{\{\u_s\}}.
\end{equation}
The map $\u_s \mapsto \u_{s+1}$ is the so-called BKL `\emph{era map}.'
Starting from $\u_0 = u_0$ it recursively determines $\u_s$,
$s=0,1,2,\ldots$, and thereby the complete Kasner sequence
$(u_l)_{l=0,1,\ldots}$.

The era map admits an interpretation in terms of continued fractions.
Applying the Kasner map to the continued fraction representation of the
initial value $\u_0$,
\begin{equation}
\u_0 =  k_0 + \cfrac{1}{k_1 + \cfrac{1}{k_2 + \dotsb}} = [k_0; k_1,k_2,k_3,\dotsc]\,,
\end{equation}
gives
\begin{align}
\begin{split}
u_0 = \mathsf{u}_0 & =
\big[ k_0; k_1, k_2,  \dotsc \big] \rightarrow  \big[ k_0 -1 ; k_1, k_2 , \dotsc \big]
\rightarrow \ldots \rightarrow \big[1 ; k_1, k_2 , \dotsc \big] \\
 \rightarrow \mathsf{u}_1 & = \big[ k_1; k_2 , k_3, \dotsc \big]
 \rightarrow  \big[  k_1-1; k_2 , k_3, \dotsc \big]
\rightarrow \ldots \rightarrow \big[1 ; k_2, k_3 , \dotsc \big] \\
 \rightarrow \mathsf{u}_2 & = \big[ k_2; k_3 , k_4, \dotsc \big]
 \rightarrow  \big[  k_2-1; k_3 , k_4, \dotsc \big]
\rightarrow \ldots\,,
\end{split}
\end{align}
and hence the era map is simply a shift to the left in the continued fraction
expansion,
\begin{equation}\label{eramap2}
  \u_s = [k_s; k_{s+1}, k_{s+2}, \dotsc] \:\mapsto\:
  \u_{s+1} = [k_{s+1}; k_{s+2}, k_{s+3},\dotsc]\:.
\end{equation}

Some of the \emph{era} and \emph{Kasner sequences\/} are periodic, notably
$u_0 = [(1)] = [1;1,1,1,\dotsc]= (1+\sqrt{5})/2$, which is the golden ratio,
gives $\u_s = (1+\sqrt{5})/2$ $\forall s$, and hence the Kasner sequence is
also a sequence with period $1$,
\begin{equation*}
(u_l)_{l\in\mathbb{N}}:\quad \sfrac{1}{2}\big(1+\sqrt{5}\big) \rightarrow
\sfrac{1}{2}\big(1+\sqrt{5}\big) \rightarrow
\sfrac{1}{2}\big(1+\sqrt{5}\big) \rightarrow
\sfrac{1}{2}\big(1+\sqrt{5}\big) \rightarrow
%\sfrac{1}{2}\big(1+\sqrt{5}\big) \rightarrow
\ldots ,
\end{equation*}
while this yields two heteroclinic cycles of period 3 in the state space
picture (since the axis permutations are not quotiented out in the state
space), see the figures in~\cite{heiugg09a}. The discretized description of
the Mixmaster chains is suitable for discussing the recent results that have
been obtained by perturbing the heteroclinic chains on $\bm{S}$, results that
can be denoted as asymptotic chain theorems, which, in contrast to the past
attractor theorems, can be regarded as `non-global state space results'
(although the two types of theorems can, of course, be combined).

%-----------------------------------------------------------------
\subsection{Asymptotic chain theorems}\label{Subsec:chaintheorems}
%-----------------------------------------------------------------

Although BKL~\cite{bkl70,bkl82,khaetal85}, as well as
Misner~\cite{mis69a,mis69b}, conjectured that the asymptotic dynamics of
Bianchi type IX is governed by the Kasner and era maps, it was only recently
that rigorous results were obtained that relate these maps to asymptotic
dynamics in Bianchi types VIII and IX. To describe these results, it is
convenient to use the following classification scheme of Kasner sequences and
associated Mixmaster chains~\cite{ugg13a,heiugg09b} (as usual the time
direction is towards the past):
%Since it has been shown that these maps, and extensions thereof, are
%associated with many interesting properties, such as chaos, it has been
%conjectured that general relativity is associated with chaotic behavior
%towards generic spacelike singularities, something that has generated
%considerable interest. However, as elaborated on in~\cite{heiugg09b}, until
%recently there were no mathematically rigorous results supporting such
%conjectures. The first proof about so-called oscillatory singularities was
%obtained by Ringstr\"om in~\cite{rin00} who showed that the singularity in
%Bianchi types VIII and IX was a curvature singularity that involved
%oscillations between at least three Kasner fix points in the
%Hubble-normalized state space picture; in a second paper~\cite{rin01}
%Ringstr\"om managed to show that the past attractor in Bianchi type IX had to
%reside on $\mathcal{A}_{\mathrm{Mixmaster}}$, but, as discussed
%in~\cite{heiugg09b}, this does not say how and if solutions are
%asymptotically related to Mixmaster chains and associated maps (and even less
%so in type VIII for which it still has not been excluded that the attractor
%also involves Bianchi type VI$_0$, see~\cite{heirin09}). However, this
%situation has changed and to describe these recent developments motivates
%revisiting a modified version of the classification of Mixmaster chains in
%terms of the Kasner parameter $u$ presented in~\cite{heiugg09b}:
%
\begin{itemize}
\item[(i)]$u_0 = [k_0; k_1, k_2, \dotsc, k_n ]$, i.e., $u_0
    \in\mathbb{Q}$. The associated Kasner sequence is finite with $n$
    eras and have an associated Mixmaster chain that terminates at one of
    the Taub points. It has been proven that these sequences are not
    asymptotically realized in the generic non-LRS case since a Taub
    point is not the $\omega$-limit set of any non-LRS
    solution~\cite{rin00,rin01,heiugg09a}.
\item[(ii)] $u_0 = [k_0;k_1,\dotsc]$ such that the sequence of partial
    quotients of its continued fraction representation is bounded, with
    or without periodicity, which corresponds to that the associated
    Mixmaster chains avoid a neighborhood of the Taub points. In the case
    of no periodicity and no cycles, B\'eguin proved that a family of
    solutions of codimension one converges to each associated
    chain~\cite{beg10}, where cycles must be excluded to avoid resonances
    in order for the proof to work. By using different techniques, and
    different differentiability conditions, Liebscher {\it et
    al\/}~\cite{lieetal10} proved explicitly that a family of solutions
    of codimension one converges to each of the $3$-cycles associated
    with $u_0 = [(1)]$. The authors also gave arguments for how their
    methods could be extended to the present general case. This was
    explicitly proved in~\cite{lieetal12}, where the authors introduced a
    new technique that involves the invariant Bianchi type I and II
    subset structure, which tie the results to the Lie contraction
    hierarchy, and hence also implicitly to basic physical principles.
\item[(iii)] $u_0 = [k_0;k_1,\dotsc]$ is an unbounded sequence of partial
    quotients, which is the generic case. The associated Kasner sequence
    is unbounded and the associated Mixmaster chain enters every
    neighborhood of the Taub points infinitely often. As argued
    in~\cite{reitru10},\footnote{Ref.~\cite{reitru10} uses different
    mathematical techniques than the other rigorous papers in this area.
    As a consequence the results, although plausible, seem to be somewhat
    controversial in the research community. Due to this, and due to an
    intrinsic value, it would be of interest if the results could be
    confirmed, or preferably even extended, with some other independent
    methods.} a subset of these chains is relevant to the description of
    the asymptotic dynamics of actual solutions: for each $u_0$ such that
    the sequence $(k_n)_{n\in\mathbb{N}}$ can be bounded by a function of
    $n$ with a prescribed growth rate, there exists an actual solution
    that converges to the chain determined by $u_0$. On the other hand,
    chains associated with initial values $u_0 = [k_0;k_1,\dotsc]$ with
    rapidly increasing partial quotients $k_n$, $n\in\mathbb{N}$ are
    perhaps less relevant for the description of the asymptotic dynamics
    of actual solutions; if a solution shadows a finite part of such a
    chain it may be thrown off course at the point where the chain enters
    a too small neighborhood of the Taub points. The prescribed bound on
    the growth rate is weak enough to yield generic continued fraction
    representations, but these results do not say anything about how many
    solutions actually converge to a given chain, nor if the asymptotic
    dynamics of a generic initial data set is represented by a
    heteroclinic chain.
\end{itemize}

The above results imply that $\mathcal{A}_{\mathrm{IX}}=\overline{\bf
B}_\mathrm{II}$ is indeed the global past attractor for Bianchi type IX, but
$\overline{\bf B}_\mathrm{II}$ is not necessarily the global past attractor
for type VIII, since it still has not been excluded that the type VIII
attractor also involves the vacuum Bianchi type VI$_0$ subset.

A spatial Fermi frame is not the only frame that is useful. Damour and
coworkers~\cite{dametal03} have used spatial Iwasawa frames to produce
"cosmological billiards" with a Chitr\'e and Misner~\cite{chi72,grav73}
procedure that relies on assumptions similar to those of BKL. Iwasawa frames
lead to a somewhat different picture of generic past dynamics than Fermi
frames, as discussed next.

%-----------------------------------------------------------------
\subsection{Transitions, concatenation and chains in an Iwasawa frame}\label{Subsec:concatIwasawa}
%-----------------------------------------------------------------

The Iwasawa approach entails diagonalizing and parameterizing the metric by a
Gram-Schmidt orthogonalization of the spatial coordinate coframe $\{d x^i\}$
(corresponding to a Cholesky decomposition of a symmetric matrix $A$ into a
product $R^T R$, where the diagonal elements of the triangular matrix $R$ are
factored out), which in turn yields a natural conformally Hubble-normalized
orthonormal frame~\cite{heietal09}, obtained by setting
\begin{equation}
E_1\!^2 = E_1\!^3 = E_2\!^3 =0 \quad \Rightarrow \quad N_3=0, \quad
(R_1,R_2,R_3) = (-\Sigma_{23},\Sigma_{31},-\Sigma_{12}).
\end{equation}
This leads to a dynamical system on $\bm{S}$ that admits a Kasner circle
$\mathrm{K}^\ocircle$ of fixed points given by setting all variables to zero
except the diagonal shear variables, which again are given by $\Sigma_\alpha
= \hat{\Sigma}_\alpha = 3p_\alpha - 1$. However, by choosing a different
frame than the Fermi frame the linearisation of $\mathrm{K}^\ocircle$ on the
dimensionless state space $\bm{S}$ on the local boundary leads to a different
result than~\eqref{geomstabeq}:
\begin{subequations}\label{Iwasawalin}
\begin{align}
\label{Adecay}
\parb_0 A_\alpha & = 3 (1 - p_\alpha) A_\alpha & & ({\rm no\,\,sum\,\, over\,\,}\alpha), \\
\label{Nalphabetadecay}
\parb_0 N_{\alpha \beta} & =
6( 1 -p_\gamma ) N_{\alpha \beta} & & (\alpha \neq \beta \neq \gamma \neq \alpha),  \\
\parb_0 N_1 & = 6 p_1 N_1, &
& \parb_0 N_{2} = 6 p_{2}\,N_{2}\, \\
\parb_0 R_1 & = 3 (p_3 - p_2) R_1, &
& \parb_0 R_2 =  3(p_3 - p_1)R_2, &
& \parb_0 R_3 = 3(p_2 - p_1)R_3,
\end{align}
\end{subequations}
where $N_1 = N_{11}$ and $N_2 = N_{22}$ (recall that $N_3 = N_{33}=0$). As in
the Fermi case, the variables $N_{\alpha\beta}$ ($\alpha \neq \beta$) and
$A_\alpha$ belong to the past stable subspace of each fixed point of
$\mathrm{K}^{\ocircle}$ (except at the Taub points). In contrast, the
variables $(R_1,R_2,R_3)$ and $(N_1, N_2)$ are stable or unstable depending
on where the point $(\hat{\Sigma}_1,\hat{\Sigma}_2,\hat{\Sigma}_3)$ is
located on $\mathrm{K}^{\ocircle}$. Finally, the variables $\Sigma_\alpha =
\Sigma_{\alpha\alpha}$ belong to the center subspace, i.e., they are
temporally constant to first order.

%We proceed by performing a local dynamical systems analysis in a neighborhood
%of the Kasner circle $\mathrm{K}^{\ocircle}$. Linearization of the dynamical
%system at an arbitrary point $(p_1,p_2,p_3)$ of $\mathrm{K}^{\ocircle}$
%yields
%\label{Edecay}
%\partial_\tau E_\alpha{}^i & = -3 (1 - p_\alpha) E_\alpha{}^i & & ({\rm no\,\,sum\,\, over\,\,}\alpha) \\
%\intertext{for the conformal frame variables, and}

As argued in~\cite{heietal09}, generically $N_{22}\rightarrow 0$ toward the
singularity, even though it is past unstable on part of
$\mathrm{K}^\ocircle$. Since $N_{33}=0$, it therefore follows that
$\mathcal{T}_{N_1}$ are the only $\mathcal{T}_{N_\alpha}$ transitions. The
$\mathcal{T}_{N_2}$ and $\mathcal{T}_{N_3}$ transitions are instead replaced
with the \emph{frame transitions} $\mathcal{T}_{R_1}$ and $\mathcal{T}_{R_3}$
(as argued in~\cite{heietal09}, generically also $R_{2}\rightarrow 0$), which
correspond to rotations of Kasner states with $\pi/2$ in the 2-3-plane and
1-2-plane, respectively; note that these rotations should \emph{not} be
confused with the BKL effect of rotation (and eventual freezing) of Kasner
axes.\footnote{Note that the arguments in~\cite{heietal09} for the
suppression of $N_2$ and $R_2$, as well as of the so-called `multiple
transitions,' were temporally non-local in character, using the heteroclinic
chains on the so-called oscillatory subsets and its associated statistical
features. Note that~\cite{heietal09} therefore does not only rely on local
stability analysis, but have in common with the more recent (rigorous)
papers~\cite{beg10,lieetal10,lieetal12} that it makes use of heteroclinic
chains.} The $\mathcal{T}_{R_1}$ and $\mathcal{T}_{R_3}$ frame transitions
correspond to spatial frame rotating versions of the `generalized' Kasner
solutions on the local boundary, with $\Sigma_1$ and $\Sigma_2$ being
temporal constants, respectively, as shown in the Bianchi type VI$_{-1/9}$
case below.

As in the case of SH class A models, transitions can be concatenated to yield
heteroclinic chains~\cite{heietal09,heietal12}. In the present case this
means that $\mathcal{T}_{N_1}$, $\mathcal{T}_{R_1}$ and $\mathcal{T}_{R_3}$
transitions form (BKL) `\emph{Iwasawa chains},' which describe the
oscillatory evolution along \emph{individual} timelines.

We refer to~\cite{heietal09} for more details for the general case. Instead
we here take a closer look at the Bianchi type VI$_{-1/9}$ vacuum models. As
stated above, the general Bianchi type VI$_{-1/9}$ models are as general as
the Bianchi type VIII and IX models and also exhibits an oscillatory
singularity (although there so far exists no formal proof of this). Since
these models can be expressed in terms of a symmetry adapted spatial Iwasawa
frame, they act as toy models for the general field equations when these are
expressed in such a frame; moreover, the generic past asymptotic behaviour of
the Bianchi type VI$_{-1/9}$ models is expected to be described by an
attractor that also describes the generic past asymptotic BKL behaviour of
the general inhomogeneous case in an Iwasawa frame~\cite{heietal09}.

%-----------------------------------------------------------------
\subsubsection*{Bianchi type VI$_{-1/9}$ models}\label{subsubsec:classB}
%-----------------------------------------------------------------

There are two invariant boundary subsets of~\eqref{VIevoleq}
and~\eqref{VIconstreq} that are of particular interest for the past dynamics:
The Bianchi type I (Kasner) subset given by $N_1=A=0$, $\Sigma^2=1$ and the
Bianchi type II subset given by $A=R_3=0$. In general both subsets are
associated with frame rotation for which $R_1^2 + R_3^2 \neq 0$, although
they both admit the standard diagonalized Fermi frame representations
described in the class A case, where the Kasner case yields the Kasner circle
of fixed points $\mathrm{K}^{\ocircle}$. Rotations of the spatial frame are
just gauge transformations whose effect can be obtained by means of frame
invariants. The first two are given by that the shear tensor is trace-free,
i.e., $\Sigma_1+\Sigma_2+\Sigma_3=0$, and that $\Sigma^2$ is determined by
the Gauss constraint, i.e., $1 - \Sigma^2 - \sfrac{1}{12}N_1^2= 1 - \Sigma^2
- \Omega_\mathrm{k} = 0$.

In the Kasner case the determinant $\det \Sigma_{\alpha\beta}$ is preserved
yielding
\begin{equation}\label{framecons}
\det \Sigma_{\alpha\beta} = \Sigma_1(\Sigma_1^2 - 3 + R_3^2) - \Sigma_3R_3^2 =
\Sigma_3(\Sigma_3^2 - 3 + R_1^2) - \Sigma_1R_1^2 = 2 + 27p_1p_2p_3,
\end{equation}
where $p_1p_2p_3 = -u^2(1+u)^2/(1+u+u^2)$ where $u$ gauge-invariantly
describes the Kasner state, which, together with the other invariants,
describes the orbits on the Kasner subset. We refer to the orbits as multiple
$\cT_{R_1R_3}$ frame transitions, since they excite more than one degree of
freedom; the special cases $R_3=0$ ($R_1=0$) gives the single $\cT_{R_1}$
($\cT_{R_3}$) frame transitions for which $\Sigma_1$ ($\Sigma_3$) is
conserved, which follows directly from~\eqref{framecons}.

In the Bianchi type II case rotations only occur in the 2-3-plane since
$R_3=0$. It follows that both $\Sigma_1$ and $N_1$ (and hence
$\Omega_\mathrm{k}$, which is more convenient to use than $N_1$ in this case)
are invariant under such transformations, and since the vacuum Bianchi type
II case only has a 2-dimensional true dimensionless state space it follows
that the equations for $\Sigma_1$ and $\Omega_\mathrm{k}$ form a dynamical
system invariant under rotations in the 2-3-plane, as can be seen explicitly
since
\begin{equation}
\parb_0\Sigma_1 = - 2\Omega_\mathrm{k}(4+\Sigma_1), \qquad
\parb_0\Omega_\mathrm{k} = -4\Omega_\mathrm{k}[\Omega_\mathrm{k} - (1+\Sigma_1)].
\end{equation}
It follows that $[(1+\Sigma_1)^2 + 3\Omega_\mathrm{k}]/(4+\Sigma_1)^2$ is a
conserved quantity that describes the orbits on the subset, which we refer to
as (multiple) $\cT_{R_1\!N_1}$ Bianchi type II transitions; the special case
of a Fermi frame $R_1=0$ yields the $\cT_{N_1}$ Bianchi type II transitions
encountered in class A. The conserved quantity can be used to reduce the
dynamical system to a $u$-parameterized 1-dimensional problem;
following~\cite{heietal09} and defining $\zeta$ according to
\begin{equation}\label{Zeq}
\Sigma_{1} = -4 + (1 + u^2)\, \zeta, \qquad
\Omega_\mathrm{k} = \frac{3}{\zeta_+ \zeta_-}\left(\zeta_+ -
\zeta\right)\left(\zeta-\zeta_-\right),
\end{equation}
where $\zeta_\pm = 3/(1 \mp u + u^2)$ and $ 0 < \zeta_- < 1$, $0 < \zeta_+ <
3$, where $u = u_-$ characterizes the initial Kasner state (time direction
towards the past), yields
\begin{equation}\label{Zeq2}
\parb_0\zeta = -2\Omega_\mathrm{k}\zeta,
\end{equation}
whose solution, by taking the asymptotic limits, gives the BKL Kasner
map~\eqref{BKLMap}. The $\cT_{N_1}$, $\cT_{R_1}$ and $\cT_{R_3}$ transitions
are associated with the towards the past unstable subsets of
${\mathrm{K}^{\ocircle}}$, which follows from restricting~\eqref{Iwasawalin}
to the present models; $N_1$ is a trigger of instability towards the past on
the Kasner arc ${\mathrm{K}_1^{\ocircle}}$; $R_1$ is a past trigger on the
arc $(132)\cup\mathrm{T}_2\cup(312)\cup\mathrm{Q}_3\cup(321)$, while $R_3$ is
a past trigger on the arc
$(321)\cup\mathrm{T}_1\cup(231)\cup\mathrm{Q}_2\cup(231)$. In contrast to the
class A case, where there were only single triggers everywhere on
${\mathrm{K}^{\ocircle}}$ (except at the Taub points), which correspond to
1-dimensional unstable subsets towards the past at each unstable fixed point
on ${\mathrm{K}^{\ocircle}}$, sectors $(321)$ and $(132)$ have two triggers,
$R_1,R_3$ and $R_1,N_1$, respectively, corresponding to 2-dimensional
unstable subsets towards the past, given by the $\cT_{R_1R_3}$ and
$\cT_{R_1N_1}$ transitions, respectively. It has been
argued~\cite{hewetal03,heietal09} that asymptotically, in some generic sense,
solutions follow the 1-dimensional subsets only, i.e., towards the past
$R_1R_3\rightarrow 0$ and $R_1N_1 \rightarrow 0$.\footnote{As discussed
in~\cite{heietal09}, the assumption that multiple transitions are generically
suppressed asymptotically is also an underlying assumption in the Iwasawa
billiard approach used by Damour {\it et al\/}~\cite{dametal03}.} Finally,
note that it is only the $\cT_{N_1}$ (and $\cT_{R_1N_1}$) transitions that
result in a change of $u$ according to the BKL Kasner map~\eqref{BKLMap}.

Based on the stability analysis of ${\mathrm{K}^{\ocircle}}$, the past
attractor of the vacuum type VI$_{-1/9}$ on $\bm{S}$ is conjectured to
satisfy
\begin{equation}\label{AVI}
\mathcal{A}_{\mathrm{VI}_{-1/9}} \in \mathrm{K}^\ocircle \cup
\mathcal{K}_{R_1\!R_3}\cup
\mathcal{B}_{R_1\!N_1},
\end{equation}
where $\mathcal{B}_{R_1\!N_1}$ is the Bianchi type II subset with $R_1N_1\neq
0$ and $\mathcal{K}_{R_1\!R_3}$ is the Kasner subset with $R_1R_3\neq 0$
(note the difference with the Bianchi type VIII and IX cases, although the
`attractor subsets' are gauge-invariantly the same, i.e., they consist of the
union of the Bianchi type I and II subsets), while the temporally non-local
analysis of~\cite{heietal09}, as well as the numerical analysis
in~\cite{hewetal03}, suggests that
\begin{equation}\label{AVI}
\mathcal{A}_{\mathrm{VI}_{-1/9}}
\in \mathrm{K}^\ocircle \cup
\mathcal{K}_{R_1}\cup\mathcal{K}_{R_3} \cup
\mathcal{B}_{N_1},
\end{equation}
or even the stronger `billiard' conjecture:
\begin{equation}\label{AVIB}
\mathcal{A}_{\mathrm{VI}_{-1/9}} = \mathrm{K}^\ocircle\cup
\mathcal{K}_{R_1}\cup\mathcal{K}_{R_3} \cup \mathcal{B}_{N_1}.
\end{equation}
Note that since the BKL Kasner map~\eqref{Kasnermap} and the associated
Kasner era map are described in terms of changes of the gauge-invariant
Kasner parameter $u$, they still give a gauge-invariant characterization of
the heteroclinic chains on e.g. $\mathrm{K}^\ocircle \cup
\mathcal{K}_{R_1}\cup\mathcal{K}_{R_3} \cup \mathcal{B}_{N_1}$, although the
chain structure looks quite different than in the SH class A case. However,
as in that latter case, it should be noted that none of the above attractor
statements say anything about how solutions asymptotically follow the
heteroclinic chains that are formed by concatenation of the various
transitions on the present representation of ${\bf B}_\mathrm{I}={\cal K}$
and ${\bf B}_\mathrm{II}$.

%-----------------------------------------------------------------
\subsection{Other spatial frame choices}\label{Subsec:framechoices}
%-----------------------------------------------------------------

Inclusion of matter generates new challenges. In that case
$\epsilon_{\alpha}\!^\beta\!_\gamma\Sigma_{\beta\delta}N^{\delta\gamma}\neq0$
in general, and hence $N^{\alpha\beta}$ and $\Sigma_{\alpha\beta}$ cannot be
be simultaneously diagonalized. In such cases the spatial Fermi and Iwasawa
frames are not the only possible useful spatial frame choices, as is
illustrated by the SH models (and hence, due to the SH/local boundary
correspondence, this suggests that other spatial frame choices might be
useful also in inhomogeneous contexts). In the SH case, the general metric of
each Bianchi type in a symmetry adapted spatial frame can be diagonalized in
a preferred manner by means of a time dependent off-diagonal (special)
automorphism transformation~\cite{jan79,jan01}. Thus, for example, the
general Bianchi type IX (VIII) metric can be diagonalized by means of the
SO(3) (SO(2,1)) group, which can be implemented by using time dependent Euler
angles~\cite{jan79,jan01} in order to diagonalize the spatial metric in a
symmetry compatible frame. In the Hamiltonian billiard picture this leads to
that the asymptotic `big billiard' in the diagonal Bianchi type IX case is
divided into six equivalent `small billiards,' to use the language
in~\cite{damlec11a,damlec11b}, obtainable from each other by means of axis
permutations. In the Hubble-normalized state space picture this corresponds
to that two of the $\cT_{N_\alpha}$ transition degrees of freedom are
replaced by two single frame transition degrees of freedom (single frame
transitions correspond to `centrifugal bounces' in Hamiltonian
approaches~\cite{jan01,dametal03} to BKL dynamics, see~\cite{heietal09}). The
situation for a small billiard is therefore completely analogous, modulo axis
permutations, to the Iwasawa case. There is however a difference. In the
Iwasawa case there is only one small billiard while there are six in the type
IX case. Thus, in the general type IX case, relevant when one for example has
a general `tilted' perfect fluid, the conjectural `BKL attractor,' projected
onto $\bm{S}$ in the Hubble-normalized picture, consists of the union of all
Iwasawa like attractors that can be obtained by means of axis
permutations.\footnote{Perfect fluids with soft equation of state lead to
asymptotic vacuum dominance, but, in the state space picture, the spatial
velocity of the fluid becomes a test field that generates `tilt' transitions,
as illustrated in~\cite{uggetal03}.} Note that since the attractor is
composed by the union of several `oscillatory attractor subsets,' this
further complicates the goal of tying initial data to specific heteroclinic
chains. Moreover, the above just describes the conjectural asymptotic
dynamics, perturbations thereof, which are necessary for investigations of
connections with e.g. heteroclinic chains, are also affected by the different
models different automorphism groups since these groups give different
expressions for `centrifugal walls,' although these expressions are all of
the same type in the asymptotic limit.

Another complication arises from the inclusion of matter. In this case there
are not only geometric automorphism structures one can make use of, but also
structures associated with the matter content. For example, one might want to
align a spatial direction with a spatial eigendirection of the stress-energy
tensor. However, in general such a direction is not compatible with the
automorphism structure, and one is forced to make a choice for what one
thinks is most important for the issue one has chosen to address. Nor is the
connection between different choices in general a simple one, instead
different choices are related by PDEs, even in the SH case.

In the general case without symmetries, there are really no preferred spatial
frame choices from a local BKL perspective, however, global issues might make
a spatial frame globally preferred. Nevertheless, note that all choices seem
to share the same gauge-invariant description of asymptotic BKL dynamics in
terms of the gauge-invariant representation of the past attractor as the
union of the (gauge-invariant) Bianchi type I and II vacuum subsets on the
local boundary (assuming asymptotic vacuum dominance) and the associated
changes in the gauge-invariant generalized Kasner parameter $u$.

%-----------------------------------------------------------------
\subsection{Comments on BKL in the conformally Hubble-normalized state space picture}\label{Subsec:BKLlocbound}
%-----------------------------------------------------------------

The BKL analysis is based on a synchronous coordinate frame and an ad hoc
procedure for producing the generalized Kasner metric as its starting point.
This metric is subsequently used to identify Kasner instabilities that are
associated with terms that involve \emph{spatial coordinate derivatives}.
However, this is purely due to the use of a coordinate frame. One of many
advantages of the present approach, is that all BKL behaviour is associated
with the local boundary (and perturbations thereof) for which all conformally
Hubble-normalized \emph{spatial frame derivatives are zero}. Moreover, the
local boundary provides the natural link between `generalized' solutions and
their connection with the SH case via the SH/local boundary correspondence.
As a consequence it is not only possible to recover the BKL results in the
present formalism, one can tie them to rigorous concepts and developments in
the SH case, and one can go beyond the BKL picture, which further shed light
on it.\footnote{Above we have shown and outlined how many of the heuristic
BKL results fit and can be derived in the present framework. However, details
concerning e.g. rotating and freezing Kasner axes have been left to the
reader; either one can use the present formalism to derive these results
directly by perturbing the generalized Bianchi type II solutions, or one can
simply translate the BKL results to it.}

The present Hubble-normalized state space picture permit us, with hindsight,
to assess the remarkable results of BKL. They basically consist of two
pieces: (i) `Local state space' results that correspond to perturbations of
$\mathrm{K}^{\ocircle}$ and the Bianchi type II solution on the dimensionless
local boundary state space $\bm{S}$. (ii) The map~\eqref{BKLMap} and
properties associated with its iteration.\footnote{In addition BKL have
numerous `local' results concerning various matter sources, using similar
`local state space techniques' as in the vacuum case, but since this is not
the focus in this paper we refrain from discussing those results.}

To do justice to the recent developments in SH cosmology, it should be
pointed out that they involve much more that just contextualizing, shedding
light and bringing rigor to BKL related issues; some of the recent
developments reveal problems and mechanisms that BKL never even discussed,
due to that the BKL analysis is limited to a heuristic local state space
analysis and impressive intuitive insights. Moreover, as has been discussed,
underlying reasons for \emph{why} BKL-like behaviour occurs at all are being
revealed. Nevertheless, much work remains.

A clarification and extension of the BKL picture requires: %
\begin{itemize}
\item[(i)] Identification of the past attractor \emph{on} the local
    boundary (since this has not yet been accomplished for e.g. Bianchi
    type VIII, this is still an open issue), \emph{and} the relationship
    between its detailed heteroclinic chain structure and past asymptotic
    dynamics (as discussed previously, there are still many open issues
    tying e.g. initial data to specific heteroclinic chains).
\item[(ii)] Perturbations of the past attractor heteroclinic chain
    structure on the local boundary into the physical state space.
\item[(iii)] Contextualization of (i) and (ii) in terms of the full
    conformally Hubble-normalized state space picture, which e.g.
    requires also taking into account the non-local, and hence `non-BKL,'
    recurring spikes, discussed next.
\end{itemize}
%

%The coupled system of equations on the local boundary and in the SH case form
%the same dynamical system, and hence the local boundary provides a new `BKL
%context' for SH dynamics.

%which give rigour \emph{On} the local boundary the heuristic BKL results can
%be tied and extended to rigorous mathematical concepts such as past unstable
%subsets, heteroclinic orbits and chains, attractors, and the use of the whole
%machinery of dynamical systems, and via the SH/local boundary correspondence
%also provide a link to the ongoing dramatic rigorous developments in SH
%cosmology. Moreover, the present Hubble-normalized state space picture do not
%only provide a context for the BKL results and their precise formulations,
%the asymptotic state space regularization also allows for numerical
%explorations.

%%%%%%%%%%%%%%%%%%%%%%%%%%%%%%%%%%%%%%%%%%%%%%%%%%%%%%%%%%%%%%%%%%
\section{Beyond BKL: Recurring spikes}\label{Sec:spikes}
%%%%%%%%%%%%%%%%%%%%%%%%%%%%%%%%%%%%%%%%%%%%%%%%%%%%%%%%%%%%%%%%%%

The simplest inhomogeneous vacuum models that (conjecturally) admit
oscillatory singularities are the general $G_2$ models, which conveniently
can be expressed in terms of a Hubble-normalized spatial Iwasawa frame. Being
the simplest inhomogeneous models with oscillatory singularities, they form
the natural testing ground for BKL locality, as well as for breaking BKL
locality. Furthermore, the $G_2$ models also occur naturally in another
context, namely solution generating techniques.

%-----------------------------------------------------------------
\subsection{Hierarchical solution generating structures}\label{Subsec:hiersolgen}
%-----------------------------------------------------------------

There are several solution generating algorithms for models with two
commuting Killing vectors. In 2001 Rendall and Weaver developed and applied
one of these techniques to asymptotic expansions~\cite{renwea01}, obtained by
means of Fuchsian methods~\cite{kicren98,ren00}, in $T^3$ Gowdy vacuum
models, i.e., OT $G_2$ vacuum models with spatial $T^3$ topology. They found
that the $T^3$ Gowdy models exhibited both `true' and `false spikes', where
false spikes where shown to be gauge artifacts while true spikes corresponded
to gauge-invariant asymptotic non-uniformities, not explainable in the BKL
picture, even though the evolution of each spatial point approaches
$\mathrm{K}^\ocircle$ on the local boundary in the present framework.

In the conformally Hubble-normalized Iwasawa based formalism, the solution
generating technique of Rendall and Weaver corresponds to alternatively
performing certain frame rotations and so-called Gowdy-to-Ernst
transformations to OT $G_2$ models in a foliation in which the area of the
symmetry orbits is purely time dependent, the so-called timelike area
gauge~\cite{elsetal02}. In 2008 Lim applied the solution generating algorithm
of Rendall and Weaver to explicit solutions instead of asymptotic expansions,
using the 1-parameter family of Kasner solutions as the initial seed
solutions~\cite{lim08}. This leads to an infinite sequence of 1-parameter
solutions that contains the Bianchi type II solutions; their frame rotated
version, which is an example of `false spike' solutions, and the 1-parameter
family of \emph{inhomogeneous} `\emph{spike solutions},' which are
expressible in terms of elementary functions. It turns out that the solutions
obtained in this way, combined with axis permutations, form the building
blocks for local BKL and non-local `non-BKL' oscillatory behaviour, and hence
\emph{the building blocks for all known generic oscillatory behaviour are
linked hierarchically to each other by means of a solution generating
algorithm}, and by means of the symmetry based subset/local boundary
correspondence.
% (no doubt this picture can be achieved by using other solution
%generating algorithms as well).

%-----------------------------------------------------------------
\subsection{Concatenation and permanent spikes in the OT case}\label{Subsec:OT}
%-----------------------------------------------------------------

The OT $G_2$ models in the timelike area gauge~\cite{elsetal02}, which, e.g.,
contain the class A Bianchi type VII$_0$, VI$_0$, II and I models, have
non-oscillatory past singularities due to that $\mathrm{K}^\ocircle$ for each
of these models admit a past stable arc. However, it is more convenient to
treat the $G_2$ models in an Iwasawa frame rather than in a Fermi frame,
especially since this gives particularly simple expressions for the OT spike
solutions. In the spatially Hubble-normalized context these
\emph{inhomogeneous} explicit solutions are described by a 1-parameter family
of trajectories, one for each value of $|x^3|$, since the solutions admit a
discrete symmetry associated with changing the sign of the spatial coordinate
$x^3$. Furthermore, the value $x^3=0$ is associated with that $N_1$ goes
through a zero (it is of course possible to make a translation and locate the
`spike surface' at any value of $x^3$), which leads to that the
Hubble-normalized spatial frame derivatives obtain $O(1)$ amplitudes for the
spike solutions and that they therefore are of similar size as the
Hubble-normalized variables. Hence, the Hubble-normalized state space is not
only asymptotically bounded and regular on the local boundary towards the
past, but also for the spiky non-local behaviour described by the spike
solutions.

The spike solutions come in two kinds: the so-called high and low velocity
solutions. The trajectories of the high velocity solutions all originate from
a common point on $\mathrm{K}^\ocircle$ and end at a different common point
on $\mathrm{K}^\ocircle$ on $\bm{S}$. For this reason, the high velocity
solutions were referred to as \emph{high velocity spike transitions}, $\Thi$,
in~\cite{heietal12}. The trajectories of the low velocity solutions also all
originate from a common point on $\mathrm{K}^\ocircle$ (as usual the time
direction is towards the past), but all `non-spike trajectories' (for which
$|N_1|>0$) end at a common point on $\mathrm{K}^\ocircle$ which differs from
that of the spike trajectories (those that correspond to a value $x^3$ for
which $N_1=0$). This gives rise to `\emph{permanent spikes}', i.e.,
non-uniformities in e.g. the curvature scalars, which cannot be explained by
the BKL picture, even though the evolution along all timelines end at
$\mathrm{K}^\ocircle$ on the local boundary. This reveals an implicit
assumption in the BKL picture, asymptotic differentiability; if the BKL
picture is to hold for an open set of timelines, avoiding non-local `spiky
features,' then this requires $|N_1|
> 0$ in an Iwasawa frame ($|N_1N_2N_3| > 0$ in a Fermi frame) for those
timelines.\footnote{There are other non-uniformities that arise from
so-called Bianchi type II spiky features, but since these can be viewed as
part of low velocity spike solutions we will refrain from discussing them
here; there are also false spike solutions associated with zeroes of $R_1$ or
$R_3$, which we likewise refrain from discussing. Instead we refer
to~\cite{heietal12} and references therein for further details.}

When it comes to the spike solutions, each solution can be viewed as
describing the evolution along an $|x^3|$-parameterized family of timelines,
which complicates the issue of concatenation (joining solutions to each other
based on their asymptotic properties towards the past and future). However,
in the case that the evolution along \emph{all} timelines of a solution share
an asymptotic limit, in the present case a fixed point on
$\mathrm{K}^\ocircle$, then one can proceed in the same manner as when
concatenating heteroclinic orbits to heteroclinic chains and match solutions
to each other at these Kasner points, thus yielding a concatenated chain of
solutions. High velocity spike transitions $\Thi$ and individual frame
transitions $\mathcal{T}_{R_3}$ have this feature and can therefore be
alternately concatenated by identifying the `final' Kasner point of one
transition with the `initial' Kasner point of another transition, see the
figures in~\cite{heietal12}. Note that this is possible because the entire
family of curves representing a $\Thi$ transition converges to a point on
$\mathrm{K}^\ocircle$ towards the future and to another point on
$\mathrm{K}^\ocircle$ towards the past. Alternate concatenation of high
velocity spike transitions $\Thi$ and individual frame transitions
$\mathcal{T}_{R_3}$ towards the past singularity yields finite chains, which
were called \textit{high velocity chains} in~\cite{heietal12}, however,
concatenation away from the singularity, yields infinite high velocity chains
that converge to the Taub point $\mathrm{T}_3$.

Due to the absence of $R_1$ in the OT models, the sector $(312)$ becomes past
stable. As a consequence heteroclinic chains on the local boundary end at
this sector, a situation that is similar to that of the Bianchi type VI$_0$
and VII$_0$ models in a diagonalized Fermi frame where one of the trigger
variables $N_\alpha$ is missing, which leads to a stable arc on
$\mathrm{K}^\ocircle$. Eventually a high velocity chain either end at the
sector $(312)$, or it is joined with a low velocity solution (or a Bianchi
type II spiky feature, the latter half of a low velocity spike solution),
which is possible since these solutions can be joined at a common Kasner
point. This yields a final non-uniform state towards the past with the
evolution along non-spike timelines ending at sector $(312)$ while `spike
timelines,' associated with $N_1=0$, end at sector $(132)$, as depicted in
fig. 8 in~\cite{heietal12}. It should be pointed out that not all OT
solutions end in this way, e.g., some high velocity solutions end at sectors
$(231)$, $(213)$, $(321)$ towards the past. However, as can be expected, a
local analysis of $\mathrm{K}^\ocircle$ shows that this requires fine tuning
since $R_3\neq 0$ in general, which leads to a subsequent $\mathcal{T}_{R_3}$
frame transition (note that zero values of triggers do not in general lead to
invariant subsets in the inhomogeneous case, in contrast to the situation in
the SH case). Thus the past evolution along each timeline in the OT case have
a specific limit associated with $\mathrm{K}^\ocircle$, and hence these
models can be regarded as being past asymptotically pointwise self-similar,
even though some of them exhibit non-uniform features not explainable by the
BKL picture.

An impressive array of mathematically rigorous results have been accomplished
as regards the $T^3$ Gowdy vacuum models, see~\cite{rin10} and references
therein. Not unexpectedly, a lot of terminology therefore exists to describe
past asymptotic behaviour for these models. However, the success of that
terminology is based upon that the evolution along each timeline end at
$\mathrm{K}^\ocircle$ (although, unfortunately, one has historically not used
$\mathrm{K}^\ocircle$ to describe this, see~\cite{heietal12}). The situation
changes dramatically in the general case where $R_1$ is no longer identically
zero. Instead of being asymptotically pointwise self-similar, the singularity
becomes oscillatory, both as regards BKL behaviour and as regards spiky
features. As a consequence, the nomenclature and tools for describing
asymptotics in $T^3$ Gowdy vacuum models are no longer adequate, but, on the
other hand, the conformally Hubble-normalized framework is
ideal.\footnote{The name of the spike solutions is a compromise. Instead of
characterizing the solutions in terms of their past complicated asymptotic
features, the nomenclature reflects their simpler asymptotic future
behaviour, namely the Kasner point the solutions originate from towards the
past, see~\cite{heietal12}. However, it would probably have been better to
name them after properties that reflect their role for generic oscillatory
singularities.}

%-----------------------------------------------------------------
\subsection{The general $G_2$ case: Concatenation and infinite spike chains}\label{Subsec:recurrspike}
%-----------------------------------------------------------------

When expressed in an Iwasawa frame, the general $G_2$ case have the same
Kasner and ${\bf B}_\mathrm{II}$ subsets on $\bm{S}$ as the Bianchi type
VI$_{-1/9}$ models, and we therefore expect that the asymptotic evolution
along timelines with BKL evolution is connected with the same past attractor
as in that case, and presumably the models also have the same asymptotic
connection with the heteroclinic chains on that attractor. However, in
contrast to the SH Bianchi type VI$_{-1/9}$ case, the inhomogeneous general
$G_2$ models also exhibit \emph{spike chains}, which in general are infinite,
and hence associated with \emph{infinitely recurring spikes}.

In contrast to the OT case, low velocity spike solutions do not yield
permanent spikes in the general case. The non-uniform structures in the OT
case are a result of these models special features, and are \emph{not}
describing generic features. They are a consequence of the stability induced
on $\mathrm{K}^\ocircle$ because $R_1=0$, i.e., because of a lack of a
trigger degree of freedom and an associated past unstable subset. The OT
models are therefore the analogues of special non-oscillatory SH models,
which show a few but not the key properties of the general oscillatory SH
models, although the difference is even greater due to an infinite
dimensional the state space.

The introduction of $R_1$ leads to that each low velocity spike solution is
combined and transported by means of a $\mathcal{T}_{R_1}$ frame transition
and then joined with part of a high-velocity solution to form a \textit{joint
low/high velocity spike transition}, $\Tco$, so that the one-parameter family
of curves that form a $\Tco$ transition, one for each $|x^3|$, all begin and
end at two distinct fixed points on $\mathrm{K}^\ocircle$ on $\bm{S}$. Thus
$\Tco$ form a `concatenation block' that can be joined with $\Thi$ and the
frame transitions to form spike chains. The name spike chain is appropriate
since the Hubble-normalized spatial frame derivatives have the same $O(1)$
magnitude as the Hubble-normalized variables during $\Thi$ and $\Tco$
transitions (which is to be contrasted with BKL evolution for which the
Hubble-normalized spatial frame derivatives are identically zero in the
asymptotic limit); however, the Hubble-normalized spatial frame derivatives
become negligible at $\mathrm{K}^\ocircle$, and hence spike chains yield
oscillating recurring spikes, as well as oscillating Kasner states. For
similar reasons, the spike chains, just like the BKL heteroclinic chains, are
in general infinite, and hence infinite oscillatory recurring spikes are to
be expected generically, just like BKL behavior yields infinite oscillatory
evolution.

In e.g. Bianchi types VI$_0$ and VII$_0$ solutions end at the stable Kasner
arc. Solutions can reach this stable part by means of shadowing finite
heteroclinic chains of $\cT_{N_\alpha}$ transitions, or they can `drop down'
on this stable arc more directly, due to different initial conditions.
However, in Bianchi type IX \emph{all} orbits are squeezed down onto
$\overline{\bf B}_\mathrm{II}$ on $\bm{S}$ and the associated heteroclinic
chains, and the BKL picture assumes that the same is going to happen in
Bianchi type VIII. There seems to be an analogous situation for the OT contra
the general $G_2$ case, both for BKL and spike dynamics. For example, the
Bianchi type II spiky features in the OT case, see~\cite{heietal12}, become
part of the spike chains in the general $G_2$ case, which follows from that
they are described by the latter part of the low velocity spike solutions.
Moreover, since the 1-parameter family of trajectories of $\Tco$ all begin
and end at two distinct fixed points on $\mathrm{K}^\ocircle$, it follows
that the non-uniform permanent spike mechanism in the OT case is gone in a
general setting; they are an artefact of that the evolution of the natural
concatenation block $\Tco$ has been interrupted, i.e., the type of
\emph{asymptotic non-uniformities found in OT models such as the $T^3$ Gowdy
models are consequences of `mutilated' generic structures}. Instead new types
of non-uniformities are generated in the general case.

In~\cite{heietal12} it was shown that each spike transition leads to a rapid
narrowing of the spatial size of the recurring spike, and hence an infinite
spike chain lead to a recurring spike that has zero spatial size
asymptotically, surrounded by BKL evolution since the spatial limit
$|x^3|\rightarrow\infty$ of the spike is described by heteroclinic BKL
chains. Hence timelines for which $N_1=0$ (again we refrain from discussing
asymptotic gauge features associated with that the frame variables $R_1$ or
$R_3$ go through a zero value) exhibit asymptotic oscillatory evolution
(characterized by a purely electric Weyl tensor), for which Hubble-normalized
spatial frame derivatives are essential, while a different type of
oscillation takes place for surrounding `BKL timelines' (oscillations
involving both the electric and magnetic parts of the Weyl tensor), which
hence leads to spatial non-uniformities, `asymptotic gravitational defects,'
that are very different from the permanent spike features in the OT case.

%-----------------------------------------------------------------
\subsection{The past attractor}\label{Subsec:pastatt}
%-----------------------------------------------------------------

Based on analytical and numerical results, we expect that we in the general
$G_2$ case only need to take into account the structure that is generated by
the heteroclinic chains that form the past attractor on the local boundary
and the subset that is associated with the spike chains, asymptotically
restricted to only describe the evolution along the timelines that form the
spike surfaces for which $N_1=0$, since we expect this to be the asymptotic
limit of the shrinking spatial size of generic infinitely recurring spikes.
In the general $G_2$ case in an Iwasawa frame this motivates the following
conjecture for the associated past attractor ${\cal A}_{G_2}$, describing
where generic asymptotic evolution resides along non-spike and spike
timelines:
\begin{equation}\label{DeltaSigma2s}
{\cal A}_{G_2} \in
\, \left\{\begin{array}{l}
\mathrm{K}^\ocircle \cup
\mathcal{K}_{R_1\!R_3}\cup
\mathcal{B}_{R_1\!N_1} \quad\,\,  \text{if}\quad  N_1 \neq 0\\[0.5ex]
{\cal K}\cup{\cal S}_{\Thi}\cup{\cal S}_{\Tco} \qquad\qquad  \text{if}\quad N_1 = 0
\end{array}\right.  ,
\end{equation}
or even the stronger `generalized billiard' conjecture
\begin{equation}\label{DeltaSigma2s}
{\cal A}_{G_2} =
\, \left\{\begin{array}{l}
\mathrm{K}^\ocircle \cup
\mathcal{K}_{R_1}\cup\mathcal{K}_{R_3} \cup
\mathcal{B}_{N_1} \quad  \text{if}\quad  N_1 \neq 0\\[0.5ex]
{\cal K}\cup{\cal S}_{\Thi}\cup{\cal S}_{\Tco} \qquad\qquad\:\:\,\,  \text{if}\quad N_1 = 0
\end{array}\right.  ,
\end{equation}
where ${\cal S}_{\Thi}$ and ${\cal S}_{\Tco}$ refers to the evolution along
timelines for which $N_1=0$, described by the subsets associated with
${\Thi}$ and ${\Tco}$; furthermore, it is assumed that $N_1=0$ only occurs at
isolated values of $x^3$, since $N_1=0$ for a continuous range of $x^3$ would
require non-generic fine tuning. The above also assumes that $R_1R_3\neq 0$
for all $x^3$ that are of relevance for the small spatiotemporal neighborhood
of the singularity that is under scrutiny, i.e., the above description does
not take into account `false' (i.e. gauge) recurring spikes induced by zeroes
in the frame rotation variables.

%-----------------------------------------------------------------
\subsection{Maps and statistics}\label{Subsec:spikestat}
%-----------------------------------------------------------------

The BKL picture assumes that the evolution of an open, asymptotically
differentiable, set of spatial points is attracted to the union of the Kasner
and Bianchi type II subsets on the local boundary, and that the evolution is
asymptotically described by (generic) heteroclinic chains on that subset. In
terms of the gauge-invariant Kasner parameter $u$, such heteroclinic chains,
irrespective if e.g. a Fermi frame or an Iwasawa frame is used, lead to a
discrete representation of the dynamics, given by the BKL Kasner
map~\eqref{Kasnermap} and its associated era
map~\eqref{eramap},~\eqref{eramap2}. Since $\u_s \in [k_s, k_s+1)$
(cf.~\eqref{eramap2}), the number $k_s$ describes the discrete length and the
number of Kasner epochs of era $s$. Therefore, passing on to the stochastical
interpretation of (generic) Kasner sequences of epochs, it follows that the
probability that a randomly chosen era $s$ of a Kasner sequence
$(u_l)_{l=0,1,2,\ldots}$ of epochs has length $m \in\mathbb{N}$ corresponds
to the probability that $k_s = m$, or, equivalently, to the probability that
$\u_s \in [m, m+1)$. Since the sequence $(k_0, k_1, k_2, \dotsc)$ arises as
the continued fraction expansion of $u_0$ this probability corresponds to the
probability that a randomly chosen partial quotient in the continued fraction
expansion is equal to $m$. This results in Khinchin's law~\cite{kin64}, which
states that the partial quotients of the continued fraction representation of
a generic real number are distributed like a random variable whose
probability distribution is given by
\begin{equation}\label{Khinchin}
\kh(m)   \,= \,  \log_2 \left( \frac{m+1}{m+2}\right) -\log_2 \left( \frac{m}{m+1} \right),
\end{equation}
which leads to
\begin{equation}\label{lengthBKL}
\mathrm{Probability}\big(\text{length of era} = n\big) = \khl(n) = \kh(n),
\end{equation}
see Table~\ref{tab:probcomp}.

In~\cite{lim08} and~\cite{limetal09} it was shown that $\Thi$ yields a map
between different Kasner states which is obtained by applying the BKL Kasner
map~\eqref{Kasnermap} twice. Remarkably the same result was obtained
in~\cite{heietal12} for $\Tco$. Thus a spike transition $\Thi$ or $\Tco$ that
takes place as part of a spike chain leads to a change in Kasner state, which
in terms of the gauge-invariant Kasner parameter $u$ results in the following
\emph{spike} (Kasner) \emph{map}:
\begin{equation}\label{spikemap}
u_+ =
\begin{cases}
u_- - 2 & u_- \in [3 ,\infty), \\
(u_- -2)^{-1} & u_- \in [2,3],\\
\big((u_- -1)^{-1} - 1 \big)^{-1} & u_- \in [ 3/2 ,2],\\
(u_- -1)^{-1} - 1 & u_- \in [1, 3/2].
\end{cases}
\end{equation}
Iterations of this spike map generate, from every initial value $u_0 \in
[1,\infty)$, a finite or infinite recurring spike-generated sequence of
Kasner epochs $(u_l)_{l=0,1,2,\ldots}$. This sequence can be partitioned into
recurring spike-induced eras, which we denote as \emph{spike} (Kasner)
\emph{eras}, or, for brevity, $\era$s. As in the usual case of eras an $\era$
is naturally defined by a sequence of monotonically decreasing values of $u$,
and hence $u_l$ and $u_{l+1}$ belong to the same $\era$ if $u_{l+1} = u_l -
2$. If $u_{l+1}$ arises from $u_l$ by one of the other three laws
of~\eqref{spikemap}, we speak of a change of $\era$. This is exemplified by
\begin{equation*}%\label{nonphases}
\underbrace{7.29 \rightarrow 5.29
\rightarrow 3.29 \rightarrow 1.29}_{\text{\scriptsize $\era$}}
\rightarrow \underbrace{2.45}_{\text{\scriptsize $\era$}} \rightarrow
\underbrace{2.24}_{\text{\scriptsize $\era$}} \rightarrow
\underbrace{4.16 \rightarrow 2.16}_{\text{\scriptsize $\era$}}  \rightarrow
\underbrace{6.14  \rightarrow 4.14 \rightarrow 2.14}_{\text{\scriptsize $\era$}}  \rightarrow \ldots
% 7.10 \rightarrow \ldots
\end{equation*}

Denote the initial ($=$ maximal) value of the Kasner parameter $u$ in $\era$
number $s$ (where $s = 0,1,2,\ldots$) by $\u_s$. The spike map induces an
$\era$ map $\u_s \mapsto \u_{s+1}$, which recursively determines
$(\u_s)_{s\in\mathbb{N}}$ from $\u_0 = u_0$, and thereby the complete spike
induced sequence $(u_l)_{l=0,1,2,\ldots}$ of Kasner epochs. The length of an
$\era$ $s$ is determined by the value of $\u_s$: If $\u_s \in [m,m+1)$ for
some $m\in\mathbb{N}$, then the length of the $\era$ is $m/2$, if $m$ is
even, and $(m+1)/2$, if $m$ is odd. In the stochastical context, in analogy
with~\eqref{Khinchin}, let $\khp(m)$ denote the
probability that a randomly chosen element %$\u_s$
of an $\era$ sequence $(\u_s)_{s=0,1,2,\ldots}$ lies in the interval
$[m,m+1)$. Let $(\u_s)_{s\in\mathbb{N}}$ be a generic spike-induced sequence
of $\era$s. Then, as shown in~\cite{heiugg13}, the probability that a
randomly chosen element of $(\u_s)_{s\in\mathbb{N}}$ lies in the interval
$[m,m+1)$ is given by
\begin{equation}\label{Khinchinp}
\khp(m)   \,= \,  \log_3 \left( \frac{m+2}{m+3}\right) -\log_3 \left( \frac{m}{m+1} \right).
\end{equation}
Moreover, as also shown in~\cite{heiugg13}, it follows that if
$(\u_s)_{s\in\mathbb{N}}$ is a generic spike induced sequence of $\era$s,
then the probability that a randomly chosen $\era$ in this sequence possesses
length $n$ is given by
\begin{equation}\label{lengthnonBKL}
\mathrm{Probability}\left(\text{length of $\era$} = n\right) =: \khpl(n) =
\log_3 \left( \frac{2 n+ 1}{2 n+3}\right) -\log_3 \left( \frac{2 n-1}{2 n+1} \right).
\end{equation}

It is of interest to compare some consequences of the probability
distribution~\eqref{Khinchin}, which determines the probabilities for
prescribed lengths of BKL eras in BKL sequences of Kasner epochs according to
eq.~\eqref{lengthBKL}, and the probability distribution~\eqref{Khinchinp},
which determines the probabilities for prescribed lengths of $\era$s in
spike-induced sequences of Kasner epochs according to
eq.~\eqref{lengthnonBKL}. As seen in Table~\ref{tab:probcomp}, $\era$s have
the tendency of being shorter than BKL eras. The probability that an $\era$
contains one epoch is larger than $50\%$, while the probability that an
$\era$ consists of $n > 1$ epochs is smaller than that of a BKL era.
Asymptotically, for $n \gg 1$,
\begin{alignat*}{2}
&\mathrm{Probability}\left(\text{length of era} = n\right) \,& & =
\, (\log 2)^{-1} \, n^{-2}\:\left( 1  - 2 n^{-1} + O(n^{-2}) \right) \,,\\
&\mathrm{Probability}\left(\text{length of $\era$} = n\right) \,& &= \, (\log
3)^{-1} \,n^{-2} \:\left(1 -  n^{-1} + O(n^{-2}) \right) \,,
\end{alignat*}
and hence the two probabilities are asymptotically proportional with a
proportionality factor $\log 2/\log 3$.

\begin{table}[Htp]
\begin{center}
\begin{tabular}{|ccccccccc|}
\hline Sequence   & 1 & 2 & 3 & 4 & 5 & 10 &  100 & 500 \\ \hline %& 1000 \\ \hline
era &  $41.50$ & $16.99$ & $9.31$ & $5.89$ & $4.06$ &  $1.20$
& $1.4\times10^{-2}$ & $5.7\times10^{-4}$  \\ %& $1.4*10^{-4}$ \\
$\era$ & $36.91$ & $16.60$ & $9.60$ & $6.28$ & $4.44$
& $1.39$ & $1.8 \times10^{-2}$ & $7.2\times10^{-4}$
\\ \hline
\hline Length  & 1 & 2 & 3 & 4 & 5 & 10 & 100 & 500 \\ \hline %& 1000 \\ \hline
era &  $41.50$ & $16.99$ & $9.31$ & $5.89$ & $4.06$ &  $1.20$
& $1.4\times10^{-2}$ & $5.7\times10^{-4}$  \\ %& $1.4*10^{-4}$ \\
$\era$  & $53.50$ & $15.87$ & $7.75$ & $4.61$ & $3.06$ & $0.83$
& $0.9\times10^{-2}$ & $3.6\times10^{-4}$ \\ \hline
\end{tabular}
\caption{This table first describes probabilities (in $\%$) that a randomly
chosen element of a BKL/spike-generated Kasner sequence of Kasner epochs
$(\u_s)_{s\in\mathbb{N}}$, is in the interval $[m,m+1)$, $m = 1,2,3,\ldots$,
see~\eqref{Khinchin} and~\eqref{Khinchinp}. The table then describe the
probabilities (in $\%$) that a randomly chosen era/$\era$ of a BKL/spike
generated sequence of Kasner epochs, is of a prescribed length. These results
are obtained from the probability distributions $\khl(m) =\kh(m)$ and
$\khpl(m)$, see~\eqref{lengthBKL} and~\eqref{lengthnonBKL}.}
\label{tab:probcomp}
\end{center}
\end{table}

It is also of interest to follow~\cite{heietal09,heietal12} and introduce
\emph{small and large curvature phases} (the nomenclature comes from the
properties of the curvature tensor on $\mathrm{K}^\ocircle$, where we recall
that the Taub points correspond to the flat spacetime). To do so, let
$\Upsilon > 3$ (although $\Upsilon\gg 1$ is most interesting). A small
curvature phase of a BKL or spike induced sequence of Kasner epochs
$(u_l)_{l\in\mathbb{N}}$ is defined as a connected and inextendible piece
$\mathcal{L} \subset \mathbb{N}$ such that $u_l > \Upsilon$ $\forall l\in
\mathcal{L}$. During a small curvature phase $u_l$ is thus monotonically
decreasing from a maximal value by the BKL map to a minimal value in the
interval $(\Upsilon, \Upsilon+1]$, while the spike map yields a monotonic
decrease from a maximum value to a minimal value in the interval $(\Upsilon,
\Upsilon+2]$. The complement of the concept of a small curvature phase is a
large curvature phase, which is defined as an inextendible piece of the
sequence of Kasner epochs such that $u_l \leq \Upsilon$ for all $l$.

While a small curvature phase can be viewed as an era/$\era$ that is
terminated prematurely at $\Upsilon$, a large curvature phase typically
consists of many eras/$\era$s, where small and large curvature phases occur
alternately. In the following BKL example, where the choice $\Upsilon = 4$
has been made, the large curvature phase contains two and a half eras.
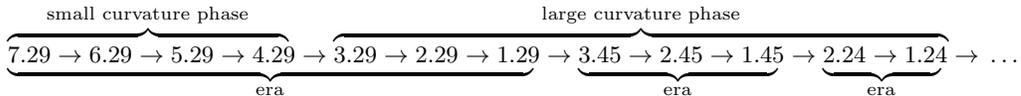
\begin{small}
\begin{equation*}%\label{scpphases}
\overunderbraces{& \br{1}{\text{\scriptsize small curvature phase}}
& & \br{5}{\text{\scriptsize large curvature phase}}}%
{& 7.29 \rightarrow 6.29 \rightarrow 5.29 \rightarrow 4.29 & \rightarrow & 3.29
\rightarrow 2.29 \rightarrow 1.29 & \rightarrow & 3.45 \rightarrow
2.45 \rightarrow 1.45 & \rightarrow & 2.24
\rightarrow 1.24 & \rightarrow  & \ldots}%
{ &\br{3}{\text{\scriptsize era}} & & \br{1}{\text{\scriptsize era}}
& & \br{1}{\text{\scriptsize era}} & }
\end{equation*}
\end{small}
Combining the probabilistic viewpoint with the concept of small/large
curvature phases lead to a fundamental result in the description of the BKL
and spike induced Kasner sequences. As shown in~\cite{heiugg13}, for generic
Kasner sequences $(u_l)_{l\in\mathbb{N}}$, \textit{small curvature phases
dominate over large curvature phases} in the following sense: Let
$(u_l)_{l\in\mathbb{N}}$ be a generic BKL or spike-induced Kasner sequence
and let $\Upsilon$ be arbitrarily large. Then for a randomly chosen epoch $u$
the probability for the event $u > \Upsilon$ is one and the probability for
the event $u \leq \Upsilon$ is zero. The underlying reason for the dominance
of small curvature phases is the failure of the probability
distributions~\eqref{Khinchin},~\eqref{Khinchinp};~\eqref{lengthBKL},~\eqref{lengthnonBKL},
to generate finite expectation values, since
\begin{equation}\label{sumdiv}
\sum\limits_{m=1}^\infty m \,\kh(m) = \sum\limits_{m=1}^\infty m \,\khl(m) = \infty ,\qquad
\sum\limits_{m=1}^\infty m \,\khp(m) = \infty =\sum\limits_{m=1}^\infty m \,\khpl(m) ,
\end{equation}
which is due to the infinite tail of the distributions. Accordingly, the
average length of an era/$\era$ is ill-defined.

%-----------------------------------------------------------------
\subsection{More about spikes}\label{Subsec:morespike}
%-----------------------------------------------------------------

A repeated application of the solution generating
algorithm~\cite{renwea01,lim08} yields solutions with an increasing number of
spikes, and applying it infinitely many times presumably lead to solutions
with inifinitely many spikes. All these solutions belong to the OT case.
However, in the general $G_2$ case the single spikes of a multiple spike
solution asymptotically become part of spike chains described by the
individual spike solutions (and the frame transitions). The reason for this
is that the different spikes in a multiple spike solution asymptotically
become causally disconnected due to asymptotic silence,\footnote{Assuming
that the heteroclinic BKL chains and spike chains are both `dominated' by the
non-Taub Kasner states that they generate on $\mathrm{K}^\ocircle$, and for
which $E_\alpha{}^i\rightarrow 0$, it seems plausible that
$E_\alpha{}^i\rightarrow 0$ in both cases, which leads to an associated
formation of shrinking particle horizons along both BKL and spike timelines,
a conclusion that also has numerical support in the case of $G_2$ models.}
and as a consequence they can individually be described by the single spike
solutions towards the singularity~\cite{limetal09}. Thus the single spike
solutions, and the associated spike chains, suffice to describe asymptotic
spike evolution in the general $G_2$ case.

Due to the Hubble-normalized state space structure that has been revealed by
the solution generating algorithm, and due to numerical investigations, it
seems reasonable to assume that there exist solutions in the inhomogeneous OT
case without spikes, with a finite number of spikes, and with infinitely many
spikes. These solutions are connected with BKL and spike chains in the
general $G_2$ case, which suggests that there in some sense might exist: (i)
a generic set of solutions without spikes and pure BKL behavior, (ii) a
generic set of solutions with a finite set of recurring spikes, (iii) a
generic set of solutions with infinitely many recurring spikes, possibly
forming a dense set (a first step to prove this would be to investigate if
the solution generating algorithm, when combined with e.g. scalings in $x^3$,
can lead to solutions with a dense set of permanent spikes in the OT case).

The presently discussed recurring spikes are located at fixed spatial
locations due to the choice of initial data. In general, recurring spikes are
moving in space. It may be that they asymptotically freeze, but this is an
open issue. If they do freeze, then our present knowledge about recurring
spikes form the first step in understanding general recurring spike behavior,
otherwise perhaps not.

It should also be pointed out that the analytic results obtained from the
solution generating technique, as well as the numerical explorations that
have been undertaken so far, all involve special initial data that excludes
or explicitly includes spikes. It is not known how spikes might form in a
dynamical situation, nor is it known if spikes can annihilate each other in
such situations, or if they can trigger additional spike formation, i.e., it
is an open issue if destructive or constructive \emph{spike interference},
respectively, can occur. It is also unknown if there are boundary conditions
associated with special physical conditions that explain the existence of
recurring spikes.

%-----------------------------------------------------------------
\subsection{The general inhomogeneous case}\label{Subsec:geninhom}
%-----------------------------------------------------------------

The primary importance of the $G_2$ models regarding generic singularities is
not the models themselves, but that they appear in the context of the
partially local $G_2$ boundary subset. Although there is a one-to-one
correspondence with the $G_2$ solutions and those on the partially local
$G_2$ boundary, the interpretation of the solutions on this latter subset is
quite different. In contrast to the $G_2$ case for which the symmetries
impose a type of `stiffness' to the spike surfaces, these surfaces are more
`flexible' in the general case without symmetries. As a consequence, spike
surfaces can a priori intersect in curves that in turn can intersect at
points in the general case, which possibly lead to new spike dynamics. At
present it is not known whether such intersections persist or recur, although
weak numerical evidence suggests that intersections only occur momentarily.
If this is correct, it follows that spike intersections may be irrelevant
asymptotically, at least in some generic sense. This would then imply that
the BKL picture in combination with $G_2$ spike oscillations may capture the
essential features of generic spacelike singularities~\cite{heietal12}.

%%%%%%%%%%%%%%%%%%%%%%%%%%%%%%%%%%%%%%%%%%%%%%%%%%%%%%%%%%%%%%%%%%
\section{Discussion}\label{Sec:disc}
%%%%%%%%%%%%%%%%%%%%%%%%%%%%%%%%%%%%%%%%%%%%%%%%%%%%%%%%%%%%%%%%%%

%-----------------------------------------------------------------
\subsection{The physical context of generic spacelike singularities}\label{Subsec:context}
%-----------------------------------------------------------------

Generic spacelike singularities are traditionally referred to as being
cosmological singularities, but it is not clear that this is their natural
physical interpretation, although one can give the following argument:  The
Big Bang singularity must have been a generic one, and presumably also a
spacelike singularity, unless the pre-inflationary universe was fine-tuned,
which in turn would require an explanation.\footnote{A multiverse scenario
would also provide possible arguments for that cosmological singularities
might be generic in nature.} On the other hand, there are several arguments
that suggest that generic spacelike singularities describe (part of) the
singularities that form due to black hole formation. Due to the complexity of
the real universe, one expects that an open set of initial data lead to black
hole formation. Seen from the outside the black holes gravitationally radiate
away their individual properties leading to \emph{external} asymptotic states
described by mass and angular momentum only (charge also remains, but in the
real universe this is unimportant), but the spacetime \emph{inside} a
physical, non-idealized, black hole is both unique and complex, where
increasing gravity eventually lead to a generic singularity that reflects the
black hole's formation. This scenario obtains further support from the
structure we have obtained for generic spacelike singularities. As previously
discussed, generic singularities are dominated by small curvature phases,
both as regards BKL behaviour and recurring spikes, i.e., they are dominated
by a Taub state. The Taub subset is in turn highly symmetric, in particular
it is axially symmetric. One might hence ask, why would a cosmological
singularity be dominated by almost axisymmetry? On the other hand, this is
precisely what one would expect for a black hole singularity.

Another interesting clue for the physical nature of generic singularities
comes from the following observation. Consider the SH Kantowski-Sachs region
of the Schwarzschild solution in the usual Schwarzschild coordinates, where
$r$ then is a timelike coordinate. Then the spacelike singularity is
asymptotically self-similar and described by the locally rotationally
symmetric state given by $u=1$, i.e., the singularity corresponds to a
$\mathrm{Q}_\alpha$ state, while the horizon corresponds to a coordinate
singularity described by a Taub state $u=\infty$. As seen from the stability
properties of $\mathrm{K}^\ocircle$, the Schwarzschild singularity is
unstable and so is the horizon structure, and due to that it is connected
with the Taub state, in a highly nonlinear fashion. The Taub points turn out
to belong to a larger Taub subset on and off the local boundary. When
considering properties of explicit solutions, this subset is connected with
horizons, caustics, and e.g. (weak) null singularities, see~\cite{limetal06}.
Due to their `Taub dominance', are generic spacelike singularities therefore
in some vague sense almost weak null singularities? If so, speculations that
black hole singularities might have a generic null singularity connected with
a generic spacelike singularity therefore gains support, from the explicit
solution examples in~\cite{limetal06} as well as from the structure of
generic spacelike singularities.\footnote{The use of solution generating
algorithms is far from exhausted when it comes to generic singularities, and
perhaps they could shed light on the possible connection between generic
spacelike and null singularities; at least it seems to be an area that has
not been explored.} Moreover, gravitational collapse may provide an
explanation for \emph{why} spikes occur; perhaps they are consequences of
increasingly nonlinear interactions of gravitational waves due to increasing
amplitudes during the final stages of gravitational collapse (this is to be
contrasted with the cosmological context: What physical motivation can
cosmology give for the structure of generic singularities, including
recurring spike formation?). Furthermore, shifting focus from cosmology to
black hole formation brings \emph{weak} cosmic censorship into light in an
area that has been dominated by closed spatial topologies and strong cosmic
censorship. The above is admittedly speculative, but so is the statement that
generic singularities are cosmological in nature. Can we afford to not take
the possibility that generic spacelike singularities might have to do with
gravitational collapse seriously?

%-----------------------------------------------------------------
\subsection{The dangers and possibilities of special models}\label{Subsec:dangerposs}
%-----------------------------------------------------------------

Much recent mathematically rigorous work on singularities has been focused on
establishing strong cosmic censorship within the context of models with quite
high symmetry and closed spatial topology, such as e.g., the $T^3$ Gowdy
models. However, as discussed in this paper, these models, which exhibit
pointwise asymptotic self-similarity, are highly misleading for generic
singularities, and it is really a generic context that is of relevance for
cosmic censorship. Moreover, a generic context puts topological issues in a
completely different light than when one deals with highly special models.
The relevance of special models for the generic cases is not the models
themselves, but their one-to-one correspondence with the local and partially
local boundary subsets, and these subsets are located in a state space
picture for generic models for which the topological issues are completely
different than for the symmetric models that are associated with these
boundaries. Furthermore, there are only a few solutions on these subsets that
are relevant in a general context, and their role is as
\emph{spatiotemporally local} (e.g. connected with the particle horizon
scale) \emph{building blocks} for the asymptotic description of generic
singularities, and hence their topology is completely irrelevant. This is
illustrated by the following: vacuum Bianchi type~I and~II models are
essential building blocks for the understanding of the singularities of the
most general Bianchi models such as Bianchi type IX, irrespective of the fact
that e.g. the Bianchi type IX models have a spatial topology that is
completely different from the spatial topologies Bianchi types I and II might
exhibit.

The key thing is hence one of `topological timing', i.e., when to address
topological issues. At this stage of affairs, there are a number of well
formulated mathematical issues that probably are of more pressing importance
than topological questions. To eventually address ultimate goals such as
cosmic censorship, one first needs to solve several proximate goals, e.g.,
the asymptotic construction of a generic spacetime in a small spatiotemporal
vicinity (i.e., in a small domain of dependence) of a generic singularity.
The present Hubble-normalized state space picture offers a way to start
addressing this issue by considering initial data in the state space that are
close to the expected attractor (`small initial attractor data' that leads to
a small spatiotemporal domain of dependence region); first within the context
of generic $G_2$ models, then for models with one or none spatial Killing
vectors.

%Moreover, the focus on global topological aspects in order to obtain a
%context for obtaining mathematically rigorous results is really quite
%unhelpful, or even detrimental, at this stage of developments.
%Are special models like the Gowdy models more misleading than helpful? The
%importance of topological "timing".

Thus it is clear that focusing too much on properties such as the topology of
special models might sometimes be detrimental for the progress of some issues
(although it might be useful for other contexts) such as the character of
generic singularities and cosmic censorship. However, some models offer the
possibility of providing manageable problems that might shed light on some
issues that are of relevance for quite general circumstances. In particular
one can construct a hierarchy of toy models that mimic various properties of
subsets that are related to the attractor for generic spacelike
singularities, in different spatial frame representations, thus providing a
mathematically \emph{and} physically progressive research program. Examples
of such toy models can be obtained from various `billiard problems,'
see~\cite{chi72,grav73,dametal03,heietal09}. Toy models are also of interest
as an intermediate step for an attempt to quantize gravity, and their
structures, especially when combined with structures revealed by solution
generating algorithms, may also provide clues and asymptotic observables that
could be used to asymptotically quantize gravity where it really needs to be
quantized, in the extreme gravity region of generic spacetime singularities

\subsection*{Acknowledgements}
It is a pleasure to thank Mark Heinzle and Woei Chet Lim for joint work in
this area, and for the many helpful and stimulating discussions that have
made this paper possible.

\end{document}